# The April 2023 SYM-H = –233 nT Geomagnetic Storm: A Classical Event

**Rajkumar Hajra[1][*], Bruce Tsatnam Tsurutani[2], Quanming Lu[1,3], Richard B. Horne[4], Gurbax Singh Lakhina[5], Xu Yang[6], Pierre Henri[7,8], Aimin Du[9,10], Xingliang Gao[1,3], Rongsheng Wang[1,3], and San Lu[1,3]**

[1] CAS Key Laboratory of Geospace Environment, School of Earth and Space Sciences, University of Science and Technology of China, Hefei, People's Republic of China.

[2] Retired, Pasadena, California, USA.

[3] CAS Center for Excellence in Comparative Planetology, Hefei, People's Republic of China.

[4] British Antarctic Survey, Cambridge, UK.

[5] Retired, B-701 Neel Sidhi Towers, Sector-12, Vashi, Navi Mumbai, India.

[6] Institute of Meteorology and Oceanography, National University of Defense Technology, Changsha, People's Republic of China.

[7] Laboratoire de Physique et Chimie de l'Environnement et de l'Espace (LPC2E), CNRS, Université d'Orléans, Orléans, France.

[8] Laboratoire Lagrange, Obervatoire Côte d'Azur, Université Côte d'Azur, CNRS, Nice, France.

[9] CAS Engineering Laboratory for Deep Resources Equipment and Technology, Institute of Geology and Geophysics, Chinese Academy of Sciences, Beijing, People's Republic of China.

[10] College of Earth and Planetary Sciences, Chinese Academy of Sciences, Beijing, People's Republic of China.

[*] Corresponding author: Rajkumar Hajra (rajkumarhajra@yahoo.co.in, rhajra@ustc.edu.cn)

**Key Points:**

- The April 2023 double-peak geomagnetic storm is a classic event with a sub-Alfvénic region located in the middle of a magnetic cloud.

- Relativistic electrons displayed classic flux variations and dayside/nightside Joule heating was the dominant storm energy dissipation.

- The near-equatorial ionospheric plasma responded to a prompt penetration electric field.

**Abstract**

The 23–24 April 2023 double-peak (SYM-H intensities of –179 and –233 nT) intense geomagnetic storm was caused by interplanetary magnetic field southward component $B_s$ associated with an interplanetary fast-forward shock-preceded sheath ($B_s$ of 25 nT), followed by a magnetic cloud (MC) ($B_s$ of 33 nT), respectively. At the center of the MC, the plasma density exhibited an order of magnitude decrease, leading to a sub-Alfvénic solar wind interval for ~2.1 hr. Ionospheric Joule heating accounted for a significant part (~81%) of the magnetospheric energy dissipation during the storm main phase. Equal amount of Joule heating in the dayside and nightside ionosphere is consistent with the observed intense and global-scale DP2 (disturbance polar) currents during the storm main phase. The sub-Alfvénic solar wind is associated with disappearance of substorms, a sharp decrease in Joule heating dissipation, and reduction in electromagnetic ion cyclotron wave amplitude. The shock/sheath compression of the magnetosphere led to relativistic electron flux losses in the outer radiation belt between $L^* = 3.5$ and 5.5. Relativistic electron flux enhancements were detected in the lower $L^* \leq 3.5$ region during the storm main and recovery phases. Equatorial ionospheric plasma anomaly structures are found to be modulated by the prompt penetration electric fields. Around the anomaly crests, plasma density at ~470 km altitude and altitude-integrated ionospheric total electron content are found to increase by ~60% and ~80%, with ~33% and ~67% increases in their latitudinal extents compared to their quiet-time values, respectively.

**Plain Language Summary**

A fast interplanetary coronal mass ejection (ICME) and its upstream sheath caused severe disturbances in the Earth's magnetosphere during 23–24 April 2023. The sheath anti-sunward of the fast ICME shock was composed of high-density plasmas and intense magnetic fields. This was followed by a plasma density rarefaction and intense magnetic fields with a field rotation (known as a magnetic cloud). This complex interplanetary structure resulted in a double-peak geomagnetic storm, and several intense auroral substorms. Solar wind kinetic energy transferred into the magnetosphere during the geomagnetic storm caused large Joule heating in the auroral ionosphere in both dayside and nightside of Earth. Compression of the magnetosphere by the shock/sheath caused losses of relativistic-energy electrons from the outer radiation belt at the beginning of the magnetospheric event. The equatorial ionospheric anomaly structure, characterized by a low plasma region on the geomagnetic equator and plasma enhancements on both sides (~±10°) of the equator, was significantly altered during the magnetic storm. In particular, the plasma density crests were more intense and expanded in hemispherical distribution. These variations are attributed to the prompt penetration electric fields to the equatorial ionosphere, which in turn modulated the equatorial ionospheric dynamics. These results should be important for prediction and modeling of geomagnetic storms and their impacts.

**1 Introduction**

Geomagnetic storms are significant disturbances of the Earth's magnetic fields (von Humboldt, 1808; Gonzalez et al., 1994; Tsurutani et al., 2003). Transient events, originating at the Sun, and modified in the interplanetary medium, lead to the geomagnetic storms (Tsurutani et al., 1988). The storms strongly affect the Earth's magnetosphere-ionosphere system, which in turn affect humankind (e.g., Tsurutani et al., 2020, 2023, 2024, and references therein). In this work, we will study the solar/interplanetary phenomena causing the magnetic storm on 23–24 April 2023, and the magnetospheric and ionospheric phenomena resulting from the storm. It is the strongest storm of the current solar cycle (starting during December 2019) prior to the recent 10–11 May 2024 superstorm (Hajra et al., 2024), comparable in strength with the largest storm (with SYM-H = −234 nT on 17–18 March 2015) of previous solar cycle 24 (this paper was submitted before May 2024).

The solar wind energy input into the magnetosphere during geomagnetic storms is believed to be caused by magnetic reconnection between southwardly directed interplanetary magnetic field (IMF) component $B_s$ and the Earth's northward magnetopause magnetic field (Dungey, 1961). This has been empirically verified by Tsurutani and Meng (1972) for substorms, and by Tsurutani et al. (1988) and Echer et al. (2008) for magnetic storms. The IMF magnitudes are intrinsically intense in the magnetic cloud (MC) portion of an interplanetary coronal mass ejection (ICME; Burlaga et al., 1981; Gonzalez et al., 1998). If the ICME is "fast" relative to the upstream magnetosonic wave speed, an interplanetary fast-forward (FF) shock will form. Behind the shock, a sheath of intensified plasma density and magnetic fields is present, which is strongly related to the shock magnetosonic Mach number (Kennel et al., 1985; Tsurutani et al., 1988). Some of the sheaths are followed by MCs, characterized by a low-density, low plasma-$\beta$ ($\beta$ is the ratio of the plasma pressure to the magnetic pressure) cold plasma, and slow and smoothly rotating IMFs (Burlaga et al., 1981; Gonzalez & Tsurutani, 1987; Marubashi & Lepping, 2007). If the magnetic fields of the MCs are intensely southward and of long-duration, a magnetic storm will occur. Similarly, if the sheath magnetic fields are strongly southward and of long-duration, they will also lead to a magnetic storm. If both interplanetary regions have southward IMFs, a double main phase magnetic storm will result (Tsurutani et al., 1988; Kamide et al., 1998; Zhang et al., 2008; Meng et al., 2019).

Magnetospheric energetic particles can be both lost and accelerated/formed during magnetic storms (e.g., Baker et al., 1994; Onsager et al., 2002; Reeves et al., 2003; Horne et al., 2005, 2009; Turner et al., 2013, 2014; Wygant et al., 2013; Hudson et al., 2014; Hajra & Tsurutani, 2018a; Hajra et al., 2020; Lu et al., 2022, and references therein). When a high plasma density sheath compresses the magnetosphere, the dayside magnetic fields become blunter than a dipole configuration. Consequently, trapped magnetospheric energetic particles can drift to the magnetopause due to the enhanced dayside magnetic field gradients. These particles will be lost to the magnetosheath and the field lines to which the energetic particles are attached will be swept away downtail by the solar wind. This process has been called magnetopause shadowing (West et al., 1972, 1981; Li et al., 1997; Kim et al., 2008; Ohtani et al., 2009; Hietala et al., 2014; Hudson et al., 2014). A second loss process can also occur. Compression of the dayside magnetosphere will cause betatron acceleration of preexisting magnetospheric energetic particles. The increase in the perpendicular kinetic energy of these particles will lead to a temperature anisotropy, which in turn will excite a plasma instability, leading to electromagnetic ion cyclotron (EMIC) wave growth, cyclotron resonant pitch angle scattering and loss of the particles to the ionosphere (Thorne & Kennel, 1971; Horne & Thorne, 1998; Summers et al., 1998; Lu et al., 2006; Meredith et al., 2006; Remya et al., 2015; Tsurutani et al., 2016; Kang et al., 2021).

During magnetic storms and substorms, ~100 eV to 1 keV plasmasheet plasma is injected deep into the nightside magnetosphere by magnetic reconnection/convection electric fields (DeForest & McIlwain, 1971). The stronger the IMF $B_s$, the stronger the magnetic reconnection at the dayside magnetopause and the stronger the nightside convection electric fields. The stronger the electric fields, the deeper the penetration of the energetic particles. As the particles are convected into higher and higher magnetic field regions, by conservation of the first and second adiabatic invariants, the particles can gain kinetic energies up to ~300–500 keV.

A somewhat rare occasion is when the solar wind becomes sub-Alfvénic near Earth (Gosling et al., 1982; Usmanov et al., 2000, 2005; Zhou et al., 2000; Fairfield et al., 2001; Smith et al., 2001; Chané et al., 2012, 2015; Lugaz et al., 2016; Hajra & Tsurutani, 2022). Rather than magnetospheric compression, the magnetosphere expands. Many of the above phenomena for magnetospheric compression will be reversed during a sub-Alfvénic solar wind interval. Energetic magnetospheric particles will be de-energized in their perpendicular components. Plasma instabilities will be quenched. Particles will drift inward/toward the Earth instead of outward, etc. (see Hajra & Tsurutani, 2022, for a statistical study on near-Earth sub-Alfvénic solar winds and their impacts).

Magnetic reconnection leads to cross tail dawn-to-dusk directed electric fields. These enhanced fields in the polar cap region can promptly penetrate to the dayside equatorial ionosphere before polar Region-2 "shielding" builds up (Tanaka & Hirao, 1973), and subsequently cause uplift of this equatorial plasma to substantially greater heights. For example, six times higher oxygen ion ($O^+$) densities at ~850 km were detected during the 30 October 2003 magnetic storm (Tsurutani et al., 2012). The limit to the height depends on the intensity of the electric field and its duration (e.g., Tsurutani et al., 2004, 2007, 2008; Mannucci et al., 2005). Low-altitude satellites will suffer enhanced drag due to this uplift (e.g., Tsurutani et al., 2007, 2022; Lakhina & Tsurutani, 2017). Upper atmospheric Joule heating is another powerful uplift mechanism in the high-latitude ionosphere (see Richmond & Lu, 2000, and references therein).

In this paper, we will study the interplanetary causes of the April 2023 strong magnetic storm, and its magnetospheric and ionospheric effects. The purpose of the present work is to determine the geophysical effects as a background for future studies dedicated to their possible societal impact.

## 2 Data and Methods

Geomagnetic storm onset, strength and variation are studied using the temporal variation of the geomagnetic SYM-H index (Iyemori, 1990), considered an accurate proxy for equatorial ring current intensity (Dessler & Parker, 1959; Sckopke, 1966). The 1-min resolution SYM-H index data are obtained from the World Data Center for Geomagnetism, Kyoto, Japan (http://wdc.kugi.kyoto-u.ac.jp/). Auroral substorm activity is studied using the SML index (1-min resolution) obtained from the SuperMAG project (https://supermag.jhuapl.edu/; Gjerloev, 2009, 2012; Newell & Gjerloev, 2011a, 2011b). A sharp decrease in SML leading to a negative bay development is associated with the westward auroral electrojet current intensification during a substorm (Newell & Gjerloev, 2011a).

The near-Earth solar wind/interplanetary conditions are studied using the solar wind plasma and IMF data (1-min resolution) obtained from NASA's OMNIWeb database (https://omniweb.gsfc.nasa.gov/). These are observations made by spacecraft located in the solar wind upstream of the Earth, the data of which are shifted in time to take into account the propagation time of the solar wind from the upstream spacecraft to the Earth's bow shock nose. The time shifting must be made in order to make a direct comparison of the solar wind variations with their geomagnetic impacts. The IMF vector components are in the Geocentric Solar

Magnetospheric (GSM) coordinate system. This system has the x-axis directed towards the Sun, and the y-axis is in the $\boldsymbol{\Omega} \times \hat{\boldsymbol{x}}/|\boldsymbol{\Omega} \times \hat{\boldsymbol{x}}|$-direction, $\hat{\boldsymbol{x}}$ is the unit vector along the x-axis, $\boldsymbol{\Omega}$ is aligned with the magnetic south-pole axis of the Earth. The z-axis completes a right-hand system. The information of the coronal mass ejection (CME) is obtained from the halo CME catalog based on the observations made by the Large Angle and Spectrometric Coronagraph (LASCO) on board the Solar and Heliospheric Observatory (SOHO) (https://cdaw.gsfc.nasa.gov/CME_list/halo/halo.html).

Potential interplanetary FF shocks (Kennel et al., 1985; Tsurutani et al., 1988) are identified from sharp and simultaneous increases in the solar wind plasma speed $V_{sw}$, proton density $N_p$, ram pressure $P_{sw}$, proton temperature $T_p$, and IMF magnitude $B_0$. Using the Abraham-Shrauner (1972) mixed-mode (plasma-IMF) technique and the Rankine-Hugoniot conservation laws (Rankine, 1870; Hugoniot, 1887; Hugoniot, 1889), we computed characteristic parameters of the shocks, namely, the shock magnetosonic Mach number $M_{ms}$, the shock speed $V_{sh}$ (relative to the upstream solar wind), and the shock angle $\theta_{Bn}$ with respect to the ambient IMF (detail description of the method can be found in Smith 1985; Tsurutani & Lin 1985; Tsurutani et al., 2011a; Hajra, 2021; Hajra et al., 2016, 2020, 2023; Hajra & Tsurutani 2018b).

The variances of the IMF components are computed at time intervals of 5, 15, and 45 min (using 1-min IMFs), and nested variances are estimated from the 45-min averages of the variances (Tsurutani et al., 1982, 2011b; Hajra et al., 2013, 2017). Solar wind wave activity is also explored using the Cluster-1 measurements (Escoubet et al., 1997) obtained from the Cluster Science Archive of the European Space Agency (ESA) (https://www.cosmos.esa.int/web/csa/access). The location of the Cluster spacecraft is inferred from the Orbit Visualization Tool (OVT, http://ovt.irfu.se) that dynamically computes the expected locations of the bow shock and magnetopause boundaries from propagated upstream solar wind conditions.

The storm energy budget is studied by computing magnetospheric energy input and energy dissipation rates in the inner magnetosphere-ionosphere system. As there is no direct measurement of magnetospheric energy transfer across the magnetopause, the Perreault and Akasofu (1978) $\varepsilon$-parameter ($\sim V_{sw}B_0^2\sin^4(\theta/2)R_{CF}^2$) is used as a proxy measurement of the magnetospheric energy input rate through the process of magnetic reconnection during southward IMFs. In the $\varepsilon$-expression, $\theta$ is the IMF clock angle (between the geomagnetic field vector and the IMF vector in the equatorial plane), and $R_{CF}$ is the magnetospheric scale size, which varies depending on the solar wind conditions (Chapman & Ferraro, 1931; Monreal-MacMahon & Gonzalez, 1997; Shue & Chao, 2013). The solar wind energy injected into the magnetosphere is dissipated into the inner magnetosphere in form of ring current (RC) energy, into the auroral ionosphere as Joule heating (JH) and auroral zone energetic particle precipitation/auroral precipitation (AP). The RC energy dissipation rate is empirically estimated as: $d$SYM-H*/$dt$ + SYM-H*/$\tau$ (Akasofu, 1981), where SYM-H* is the solar wind pressure-corrected SYM-H index (Burton et al., 1975; Gonzalez et al., 1989; Turner et al., 2001), and $\tau$ is the average RC decay time, taken as 8 hr for the present study (Yokoyama & Kamide, 1997; Guo et al., 2011). The JH rate is calculated according to: $a$|PC| + $b$PC$^2$ + $c$|SYM-H| + $d$SYM-H$^2$ (Knipp et al., 2004), where PC is the polar cap potential index, and $a$, $b$, $c$, and $d$ are season-dependent constants. The AP rate is calculated as: $10^8 \times$SME (adaptation of Akasofu, 1981). SME is the SuperMAG AE index. We have also estimated the JH rates at four equally sized magnetic local time (MLT) sectors centered at 00:00 MLT, 06:00 MLT, 12:00 MLT, and 18:00 MLT. Here, 00:00 MLT corresponds to all MLTs from 21:00 MLT to 03:00 MLT, with similar 6-hr intervals applying to the other time sectors. These MLT-dependent JH estimation is based on the "partial" ring current indices or the symmetric ring current indices (SMR) partitioned by MLT, obtained from SuperMAG. It may be mentioned that the above-mentioned empirical methods of energy budget study have been used for geomagnetic storms of varying intensity, and driven by different interplanetary causes. To mention a few, Turner et al. (2009), used these methods to study energy partitioning of 91 storms (driven by corotating interaction regions) with Dst intensity of –50 to –156 nT, and 118 CME-driven storms of Dst between –51 and –489 nT; Guo et al. (2011) compared the magnetospheric energy transfer for intense geomagnetic storms with Dst between –100 and –422 nT, driven by ICMEs and their sheaths using the above methods.

We explored the field-aligned currents (FACs; Zmuda et al., 1966; Cummings & Dessler, 1967) using the northern hemispheric radial FAC measurements provided by the Active Magnetosphere and Planetary Electrodynamics Response Experiment (AMPERE; https://ampere.jhuapl.edu/). The FAC estimations are based on the magnetic perturbations measured by sixty-six satellites of the Iridium telecommunication network in six polar orbital planes at an altitude of 780 km (orbital period of 104 min) (Waters et al., 2001, 2020; Anderson et al., 2002, 2021).

The Earth's outer radiation belt is studied using the geosynchronous ($L = 6.6$) orbit ~76 keV to 2.9 MeV electron fluxes, and the $L^*$-shell variations of the > 2.0 MeV and > 0.8 MeV (Figure 7c) electron fluxes (at ~88° pitch angles). Here $L$ is a dimensionless parameter representing the maximum radial extent, in Earth radius ($R_\oplus$), of a dipole geomagnetic field (McIlwain, 1961). $L^*$ or Roederer $L$-parameter is related to the third adiabatic invariant of the particle motion in the geomagnetic field (Roederer, 1970). The geosynchronous orbit electron fluxes in several differential channels between ~76 keV and ~2.9 MeV are obtained from GOES-18 (Geostationary Operational Environment Satellite; https://www.ngdc.noaa.gov/stp/satellite/goes-r.html). The > 0.8 MeV and > 2.0 MeV electron $L^*$-shell variations are obtained from the Satellite Risk Prediction and Radiation Forecast (SaRIF) system (https://swe.ssa.esa.int/web/guest/sarif-federated) made available by ESA. These represent "re-constructed" radiation belts based on the observed near-Earth solar wind and geosynchronous orbit electron variations and geomagnetic conditions (see Horne et al., 2013, 2021; Glauert et al., 2014, for details).

Ionospheric impacts of the geomagnetic storm are studied using the latitudinal variation of ionospheric plasma density along the Swarm C satellite orbit, and the latitudinal variation of the altitude-integrated electron number or ionospheric total electron content (TEC) measured by Swarm C. The Swarm C satellite is one of the three-satellite Swarm constellation operated by ESA (Olsen et al., 2013; Knudsen et al., 2017). The Swarm satellites are in circular orbit and Swarm C was at an inclination of ~87.4°. In April 2023, Swarm C orbited Earth at ~470 km altitude and has been at 05:00 and 17:00 local times (LTs). The Swarm data are obtained from the ESA Swarm Data Access (https://swarm-diss.eo.esa.int).

## 3 Results

Figure 1h shows the temporal variation of the geomagnetic SYM-H index during 22–25 April 2023. Gradual decrease of SYM-H from a value of ~0 nT at 09:00 universal time (UT) on 23 April marks the storm main phase onset. The main phase is characterized by two major SYM-H peaks of −179 nT at 21:59 UT on 23 April, and of −233 nT at 04:03 UT on 24 April. Thus, this storm is classified as a classical double-peak intense storm (Tsurutani et al., 1988; Kamide et al., 1998). The storm recovery began after the second SYM-H peak, and SYM-H recovered to a "quiet" value of −41 nT at 22:44 UT on 24 April. The duration of the storm main phase is 19 hr and 3 min, and that of the recovery phase is 18 hr and 41 min.

Auroral substorm activity during the geomagnetic storm is studied using the SML index (Figure 1i). The substorm onsets based on the SML index are shown by histograms. Multiple intense substorms with the SML peaks of −1311 nT (~11:21 UT), −1213 nT (~16:46 UT), −1406 nT (~17:48 UT), −1608 nT (~18:45 UT), −1570 nT (~19:45 UT), −2018 nT (~20:20 UT), and −1660 nT (~21:37 UT on 23 April) are recorded during the first-step main phase development. The second and most intense SYM-H peak is associated with an intense substorm with an SML peak of −2760 nT (~03:54 UT on 24 April). The latter is defined as a supersubstorm (SSS, with SML < −2500 nT; Tsurutani et al., 2015; Hajra et al., 2016).

Figures 1a–1f show the solar wind/interplanetary conditions during the geomagnetic storm. The solar wind data represents the interplanetary counterpart of a CME or an ICME. The CME, estimated to have a mass of ~$1.5{\times}10^{16}$ g, a plasma kinetic energy of ~$1.3{\times}10^{32}$ erg, was ejected at a linear speed of ~1284 km s$^{-1}$ from the Sun (source location: 22°S latitude, 11°W longitude) at ~18:12 UT on 21 April. This was associated with an M1.7 X-ray flare, which erupted at ~17:44 UT from the active region AR3282. The following shock, sheath and ICME features leading to the geomagnetic activities are described below.

The storm main phase onset at 09:00 UT on 23 April is preceded by a southward turning of the IMF $B_z$ at ~07:54 UT (Figure 1f). This is followed by an almost steady southward IMF/$B_s$ of ~11 nT. The vertical dashed line at ~17:41 UT on 23 April marks an interplanetary FF shock impingement onto the magnetosphere. The shock is characterized by sharp and simultaneous increases in the solar wind $V_{sw}$ (~358 to ~495 km s$^{-1}$, Figure 1a), $N_p$ (~5.7 to ~16.7 cm$^{-3}$, Figure 1c), $P_{sw}$ (~1.2 to ~8.0 nPa, Figure 1c), $T_p$ (~$1.7{\times}10^4$ to ~$25.2{\times}10^4$ K, Figure 1d, black), and IMF $B_0$ (~10.3 to ~25.2 nT, Figure 1f). The shock was encountered by the Wind spacecraft at ~16:58 UT on 23 April at a geocentric distance of ~200 $R_\oplus$ upstream Earth. The shock is computed to have a magnetosonic Mach number $M_{ms}$ of ~2.7. The shock moved at a speed $V_{sh}$ of ~247 km s$^{-1}$ (relative to the upstream solar wind) at an angle $\theta_{Bn}$ of ~81° with respect to the ambient IMF (a quasi-perpendicular shock).

The (ram) pressure increase across the interplanetary shock is calculated to cause an earthward movement of the bow shock nose from a pre-shock geocentric distance $r_{BS}$ of ~14.4 to ~8.3 $R_\oplus$ following the shock/sheath impingement (Figure 1g). The magnetopause moved earthward from a geocentric magnetopause distance $r_{MP}$ of ~10.0 to ~7.2 $R_\oplus$ (Figure 1g). The bow shock and magnetopause locations are estimated by typical models, such as Farris and Russell (1994) and Shue et al. (1998), respectively. The shock causes a simultaneous sharp rise in SYM-H from −65 to −49 nT (Figure 1h) or a sudden impulse (SI$^+$) amplitude of +16 nT, and the triggering of an intense substorm with the SML peak of −1406 nT (Figure 1i).

In the interplanetary sheath (marked by a green bar at the top of Figure 1) downstream of the shock, the plasma and IMFs are strongly compressed up to ~01:17 UT on 24 April. The sheath is characterized by high $V_{sw}$ of ~752 km s$^{-1}$, $N_p$ of ~30.3 cm$^{-3}$, $P_{sw}$ of ~19.5 nPa, and $T_p$ of $\sim 31.0 \times 10^5$ K. While the IMF components were highly fluctuating in amplitude and polarity, the IMF $B_s$ was strengthened up to $\sim$25 nT at $\sim$19:13 UT on 23 April. This causes a strong interplanetary electric field $E_{sw}$ of $\sim$14 mV m$^{-1}$ (Figure 1e). The sheath $B_s$ caused the first major storm main phase development characterized by a SYM-H peak intensity of $-$179 nT at 21:59 UT on 23 April (Figure 1h), and a large number of intense substorms (Figure 1i). The northward turning of IMF at $\sim$21:03 UT (on 23 April) resulted in significant reduction in substorm occurrences.

The interplanetary sheath is followed by a slowly decreasing $V_{sw}$, and sharp decreases in $N_p$, $P_{sw}$, and $T_p$, and a sharp north-to-southward turning of IMF $B_z$. The rarefied ($N_p$ $\sim$0.2–2.6 cm$^{-3}$) and cold ($T_p$ $\sim$1.5$\times$10$^4$ K) plasma with low $\beta$ ($\sim$0.01) can be identified from $\sim$01:17 UT up to $\sim$22:11 UT on 24 April. During this interval, the IMF $B_0$ is slowly decreasing, and the IMF components exhibit a slow and smooth rotation in polarity. This represents a flux rope MC (marked by a red bar at the top and a gray shading in Figure 1). The radial extent of the MC is estimated to be $\sim$0.28 astronomical unit (AU). It has a south-to-north magnetic component configuration, with a peak $B_s$ of $\sim$33 nT at $\sim$01:21 UT on 24 April. The IMF $B_s$ is associated with an enhanced $E_{sw}$ of $\sim$21 mV m$^{-1}$. This $B_s$, through magnetic reconnection, led to the major second-step development of the storm main phase, SYM-H reaching $-$233 nT at 04:03 UT on 24 April (Figure 1h).

Thus, the double-peak intense storm is caused by a combined effect of an interplanetary sheath and an MC (however, it should be noted that not all double-peak storms are caused by these two interplanetary structures; Meng et al., 2019). The storm recovery started with weakening of the IMF $B_s$, and SYM-H recovered to its quite-time value at the end of the MC at 22:44 UT on 24 April.

Inside the MC, low $N_p$ values are associated with high Alfvén wave intrinsic speed $V_A$ (Figure 1a) and low Alfvén Mach number $M_A$ (Figure 1b). Near the center of the MC, $N_p$ sharply falls from a value of $\sim$2.7 cm$^{-3}$ at $\sim$12:00 UT to a minimum value of $\sim$0.24 cm$^{-3}$ at $\sim$14:09 UT on 24 April. This led to a sharp rise in $V_A$ (to $\sim$1110 km s$^{-1}$ at $\sim$14:08 UT), significantly higher than $V_{sw}$ ($\sim$576 km s$^{-1}$), from $\sim$12:25 UT to $\sim$14:33 UT on 24 April, when the solar wind became sub-Alfvénic with $M_A$ <1 (marked by a light-cyan shading). The minimum $M_A$ is 0.5 at $\sim$12:48 UT. During this $\sim$2.1-hr long sub-Alfvénic solar wind interval, $T_p$ exhibits an increase from $\sim$1.9$\times$10$^4$ to $\sim$5.1$\times$10$^4$ K, the bow shock presumably disappeared (the sub-Alfvénic solar wind is sub-magnetosonic as well, with a minimum magnetosonic Mach number $M_{ms}$ = 0.5 (not shown) during the interval), and magnetopause is calculated to have moved sunward, from a geocentric distance of $\sim$9.5 to 12.2 $R_\oplus$. This interval corresponds to the recovery phase of the magnetic storm. The SYM-H peak was $-$120 nT, and it changed only at a rate of $\sim$3.5 nT h$^{-1}$ across the sub-Alfvénic interval, indicating that no significant geomagnetic storm ring current enhancement was caused by the sub-Alfvénic wind (Figure 1h). These results are consistent with those of Hajra and Tsurutani (2022). Interestingly, the sub-Alfvénic interval is associated with disappearance of auroral substorms (Figure 1i).

Figure 2 shows the normalized nested variances of the IMF components: $\sigma_x^2/B_0^2$, $\sigma_y^2/B_0^2$, and $\sigma_z^2/B_0^2$. As expected, the MC is characterized by significant decreases in the IMF variances compared to those during the sheath (e.g., Hajra et al., 2022). A further local reduction in variances can be noted during the sub-Alfvénic interval (e.g., Hajra & Tsurutani, 2022). Low variances indicate reductions in wave activity (up to the Nyquist frequency of the time series) in the sub-Alfvénic solar wind (Tsurutani et al., 1982, 2011b; Hajra et al., 2013, 2017).

Before and after the sub-Alfvénic wind interval, the Cluster-1 spacecraft was located in the magnetosheath, and it remained located upstream of the magnetopause during the sub-Alfvénic wind interval, as suggested from OVT. This was also confirmed by the plasma signatures seen in the Cluster's Waves of High Frequency and Sounder for Probing of Electron Density by Relaxation experiment (WHISPER) instrument (not shown here). Note that the existence of a proper magnetosheath region during the sub-Alfvénic wind interval is unlikely to occur. Figure 3b shows the frequency–time spectrogram of magnetic field measured by the Cluster-1 spacecraft from 10:00 UT to 18:00 UT on 24 April. Red and yellow traces in the spectrogram around frequencies of 0.2 and 1 Hz indicate EMIC waves (with typical frequencies of ~0.1–5 Hz) (Cornwall, 1965; Kennel & Petschek, 1966) with amplitudes of ~0.1–0.5 nT. During the time interval when $M_A < 1$ (Figure 3a), the EMIC waves seem to disappear promptly. The power of 0.1–10 Hz waves exhibits a sharp fall from a typical value of ~1–2 $nT^2$ before the sub-Alfvénic interval to a value of ~0.2 $nT^2$ during the sub-Alfvénic interval (Figure 3c). It may be noted (from search-coil magnetometer observation, not shown here) that higher-frequency (up to ~100 Hz) waves are also significantly reduced in amplitude during the sub-Alfvénic interval.

The storm energy budget is analyzed in Figure 4. The IMF southward turning at 07:54 UT on 23 April leads to magnetic reconnection resulting in solar wind kinetic energy transfer to the magnetosphere. The $\varepsilon$-parameter clearly indicates energy transfer starting with the IMF southward turning (Figure 4d). Strengthening of $B_s$ by the FF shock impingement at 17:41 UT on 23 April caused an increase in the solar wind energy input rate with a $\varepsilon$ peak value of ~76×10[11] W at ~19:37 UT on 23 April (Figure 4d). This corresponds to the first-step decrease of the storm main phase. Correspondingly, energy dissipation rates in the magnetosphere-ionosphere system (RC, JH, and AP) are found to increase coincident with this time (Figures 4e–4g).

Following the interplanetary sheath, a sharp southward turning of IMF and a strong $B_s$ field during the MC lead to a sharp increase in $\varepsilon$ up to ~173×10[11] W. During the entire main phase of the storm (from 09:00 UT on 23 April to 04:03 UT on 24 April), the total reconnection energy input from the solar wind is estimated to be ~215×10[15] J (Figure 4h), and only ~32% of this is dissipated into the inner magnetosphere-ionosphere system in forms of JH (26%) and AP (3%) in the auroral ionosphere, and RC (3%) development in the inner magnetosphere. In particular, the major part of the total dissipated energy (~69×10[15] J) is accounted by the JH dissipation (81%), while AP (11%) and RC (8%) contributions to the total dissipation energy are minor. Presumably, the remainder of the solar wind reconnection energy is lost downtail. The sub-Alfvénic solar wind interval is characterized by a sharp fall in JH dissipation (Figure 4e).

Figure 5 shows the JH rates at four equally sized MLT sectors centered at 00:00 MLT, 06:00 MLT, 12:00 MLT, and 18:00 MLT. The temporal variations of the heating rates are found to be comparable during all four local time sectors (Figure 5b). In addition, during the storm main phase, the dayside ionosphere JH rates at 06:00 MLT and 12:00 MLT sectors are highly correlated to the nightside ionosphere heating rates at 18:00 MLT and 00:00 MLT sectors (Figures 5c–5d).

In Figure 6, we explore the northern hemispheric Birkeland/FACs rooted in the polar ionospheric *E*-region during different phases of the storm. These are shown as maps in Altitude Adjusted Corrected GeoMagnetic (AACGM) latitude (Baker & Wing, 1989) and MLT coordinate system. Figure 6c corresponds to the storm main phase onset. Signatures of weak upward Region-1 (around 70°–75° AACGM latitude) and downward Region-2 (around 65°–70° latitude) currents observed during noon to pre-midnight sector are associated with magnetic convection initiated with the IMF southward turning.

Figure 6d corresponds to an intense substorm peak (SML = –2018 nT) during the magnetic storm first main phase development under strong IMF $B_s$. The intense substorm-related DP1 (disturbance polar) currents can be observed around the 00:00 MLT sector in a large region extending from ~60° to ~75° latitudes. In addition to this, even stronger Region-1 currents extending up to ~80° latitude, and Region-2 currents extending up to ~60° latitude, are observed in almost all MLT sectors. This global-scale current system, associated with fluctuations in the magnetospheric plasma convection under strong sheath $B_s$, is called the DP2 current (Nishida, 1968). During the storm main phase second peak under even stronger MC $B_s$, both the DP1 and DP2 currents are found to be stronger (Figure 6e).

Figures 6f–1g corresponds to recovery phase of the magnetic storm, before and during sub-Alfvénic wind interval, respectively. During the sub-Alfvénic wind interval, the disappearance of the DP1 current (Figure 6g) is associated with disappearance of auroral substorms (Figure 1i). The overall DP2 current also decreased with decreasing magnetospheric convection due to northward turning of IMF.

The Earth's outer radiation belt exhibited dramatic variations during the geomagnetic storm. Figure 7 shows the geosynchronous orbit ~76 keV to 2.9 MeV electron fluxes (Figure 7a), and the $L^*$-shell variations of the > 2.0 MeV (Figure 7b) and > 0.8 MeV (Figure 7c) electron fluxes (at ~88° pitch angles). The solar wind $V_{sw}$, $V_A$ (Figure 7d), IMFs (Figure 7e), SYM-H (Figure 7f), and SML (Figure 7g), and the markings of the interplanetary structures (FF shock, sheath and MC) are repeated from Figure 1 for references. As can be seen from Figure 7a, the entire storm period is characterized by large fluctuations in ~76 keV to 1.5 MeV electron fluxes (2.0–2.9 MeV electrons do not exhibit significant variations above their noise levels) at the geosynchronous orbit. Electron injections (Figure 7a) are clearly correlated to the substorm activity or the SML enhancements (Figure 7g). The FF shock impingement at ~17:41 UT on 23 April (vertical dashed line) is associated with prompt flux decreases. The $L^*$-shell variations of the > 0.8 and > 2.0 MeV electrons (Figures 7b–7c) show that the ~MeV electrons are lost in the outer radiation belts for $L^*$ = 3.5–5.5, and the fluxes are enhanced in the lower $L^*$ region ($L^* \leq$ 3.5) after the shock/sheath impingement. As the radiation belts were transported closer to the Earth, the risk of satellite charging for satellites in GPS type orbits, which cross the equator near $L$ ~4.2, decreased during the storm, but increased for those in lower orbits. Sub-Alfvénic solar wind seems to have no significant impacts on the outer radiation belt MeV electrons.

Figure 8a shows the (geomagnetic) latitudinal variation of ionospheric plasma density along the Swarm C satellite orbit, and Figure 8b shows the (geomagnetic) latitudinal variation of TEC measured by Swarm C during 22–25 April. They correspond to the 17:00 LT passes. Variations on 22 April up to about the middle of 23 April show the quiet-time structures. These are characterized by equatorial ionization anomaly (EIA, with a density trough at the magnetic equator and two plasma crests on both sides of the equator; Appleton, 1946; Hanson & Moffett, 1966; Bailey et al., 1997). From ~00:00 UT on 22 to 12:00 UT on 23 April (geomagnetic quiet), the minimum plasma density and the minimum TEC are recorded around the magnetic equator, while the maxima are observed around ±15° magnetic latitudes. The crest-to-trough ratio is ~3 for plasma density, and ~1.5 for TEC. However, during the magnetic storm main phase, the anomaly structure is significantly modified, with overall enhancements in the plasma density and TEC. For example, at ~22:01 UT on 23 April, the northern crest of the plasma density (TEC) of ~3.5×10$^6$ cm$^{-3}$ (~100 TECU) is observed at ~20° (~25°) latitude, and the southern crest with the plasma density (TEC) of ~3.0×10$^6$ cm$^{-3}$ (~90 TECU) is observed at –20° (–25°) latitude (TEC is expressed in the TEC unit/TECU, where 1 TECU = 10$^{16}$ electrons m$^{-2}$). The anomaly trough with plasma density (TEC) of ~0.7×10$^6$ cm$^{-3}$ (40 TECU) is observed around the magnetic equator. The crest-to-trough ratio in plasma density (TEC) is ~5–5.7 (~2.3–2.8) during the storm main phase. In general, around the anomaly crests, the storm main phase plasma density exhibits a ~60% increase and TEC exhibits a ~80% increase, and their crest-to-crest latitudinal extents increased by ~33% and ~67% compared to their quiet-time values, respectively. These storm-time changes in ionospheric plasma density and TEC structures are most probably due to the prompt penetrating electric field (PPEF; Tsurutani et al., 2004, 2008).

## 4 Summary and Discussion

Major findings of this study on the 23–24 April 2023 intense geomagnetic storm are summarized and discussed below.

1. The double-peak intense geomagnetic storm is caused by an interplanetary sheath-MC interaction. The sheath IMF $B_s$ of ~25 nT and the MC $B_s$ of ~33 nT lead to two SYM-H peaks of –179 nT and –233 nT, respectively. Hajra and Tsurutani (2018a) showed that the strongest storm of the last solar cycle (24, 2008–2019) occurring on 17–18 March 2015 exhibited a three-peak SYM-H main phase. While the first SYM-H peak of –101 nT was associated with a sheath $B_s$, the second and third SYM-H peaks of –177 nT and –234 nT, respectively, were caused by the MC $B_s$ fields. A similar mechanism was suggested for another three-peak geomagnetic storm (with SYM-H peaks of –31 nT, –93 nT, and –146 nT) occurring on 7–8 September 2017 (Hajra et al., 2020). The recent superstorm of 10–11 May 2024, which is the second largest geomagnetic storm (after the March 1989 SYM-H = -720 nT storm) in the space age, was characterized by three SYM-H peaks of –183, –354 and –518 nT. Hajra et al. (2024) showed that the first phase was caused by a FF shock and a sheath $B_s$ of ~40 nT. The cause of the second phase was a deepening of the sheath $B_s$ to ~43 nT. A magnetosonic wave compressed the sheath to a $B_s$ of ~48 nT causing the third and most intense storm main phase. It may be mentioned that 67% of all intense storms (with Dst < –100 nT) occurring between 1957 and 1991 (Kamide et al., 1998), and 49% of all intense storms occurring between 1996 and 2006 (Zhang et al., 2008) are reported to be characterized by double-peak main phases. Kamide et al. (1998) suggested the roles of two distinct processes in developing double-peak geomagnetic storms, namely, 1) the magnetic reconnection-driven magnetospheric convection (Burton et al., 1975; McPherron, 1997) causing the first ring current enhancement, followed by 2) a substorm-related ring current ion accumulation (Daglis, 1997; Yue et al., 2011, 2016) during the second main phase development. Tsurutani et al. (1988) suggested that the two-stage intense storm developments are mainly associated with a combined effect of interplanetary sheath and driver gas (or MC) fields. This is consistent with the present result. However, Meng et al. (2019) suggested that superstorms (Dst $\leq$ −250 nT) with two or more Dst/SYM-H peaks may be caused by either the combined effects of the sheath $B_s$ and the MC $B_s$ or by two MCs or by a single MC with double dip $B_s$ amplitudes.

2. While the geomagnetic storm first-step main phase is associated with multiple intense substorms (SML intensity of −1000 to −2000 nT), only one SSS (SML peak = −2760 nT) occurred during the second and strongest storm main phase peak. In other words, the storm second main phase development and the SSS are essentially the same. Despirak et al. (2024) reported that these substorm activities were associated with strong (~12–46 A) geomagnetically induced currents (GICs) in the north-west Russia and south Finland regions. Tsurutani and Hajra (2023) reported several cases where an SSS is the only substorm that occurs during a magnetic storm. As proposed by Hajra et al. (2020), the O$^+$ ions injected during the SSS event (Daglis, 1997; Yue et al., 2011, 2016) may be part of the energy source of the second main phase ring current development.

3. The geomagnetic storm peak is characterized by a peak solar wind energy input rate of ~1.7×10$^{13}$ W due to magnetic reconnection under strong IMF $B_s$. A total of ~21.5×10$^{16}$ J solar wind energy is injected into the magnetosphere during the storm main phase. In the absence of direct measurements of solar wind-magnetosphere-ionosphere energy coupling, empirical estimations are used to understand the geomagnetic storm energy budget (see, Vichare et al., 2005; Turner et al., 2006, 2009; Guo et al., 2011, 2012; Hajra et al., 2014; Tsurutani & Hajra, 2023, and references therein). Thus, the energy values presented here possibly have uncertainties; however, relative values (percentages) might be useful for comparison. During the storm main phase, only ~32% of the total solar wind reconnection energy transferred to the magnetosphere is found to be dissipated in the inner magnetosphere-ionosphere system. One surprising result is that major part of total dissipation energy (~6.9×10$^{16}$ J) goes to Joule heating (81%), while ring current energy (8%) is significantly less during this intense storm main phase. This result is in clear opposition to earlier results that ICME-driven intense storms dissipate more of the transferred energy in the ring current (e.g., Monreal-MacMahon & Gonzalez, 1997; Vichare et al., 2005), while Joule heating dissipation dominates during the substorm events (e.g., Østgaard et al., 2002; Tanskanen et al., 2002; Tenfjord and Østgaard, 2013; Hajra et al. 2014). In particular, Monreal-MacMahon and Gonzalez (1997) reported that ring current (Joule heating) injection, on average, accounts for ~49% (~12%) of the solar wind energy input during the main phases of ICME-driven superstorms (Dst < −240 nT). According to Vichare et al. (2005), on average, ring current dissipation and Joule heating account for ~72% and ~19% of the total magnetospheric energy dissipation, respectively. Lower ring current injection and higher Joule heating observed during the present intense storm main phase might be due to the occurrence of multiple intense substorms in the first-step main phase development and an SSS during the second-step main phase development. Tsurutani and Hajra (2023) found that during the SSS events, the major part of the solar wind input energy is dissipated into Joule heating (~30%), with substantially less energy dissipation in auroral precipitation (~3%) and ring current energy (~2%). These numbers are consistent with the present results.

4.  Joule heating is found to be equally distributed in all local time sectors during the intense geomagnetic storm. Tsurutani and Hajra (2023) recently reported that during the SSS events, the dayside Joule heating was comparable to the nightside Joule heating, giving a picture of an isotropic global energy dissipation in the magnetospheric/ionospheric system. This recent finding indicates the possibility that Joule heating in general is a global phenomenon for strong geomagnetic disturbances bringing back the Axford and Hines (1961) viscous interaction picture. Nishida (1968) argued for a global-scale current system associated with fluctuations in the magnetospheric plasma convection. This DP2 current is different from the substorm-related DP1 current system. We observed strong DP2 currents during the storm main phase, consistent with global Joule heating distribution.

5.  The geomagnetic storm main phase is characterized by ~76 keV to 1.5 MeV electron injections at the geosynchronous orbit, correlated to the substorm SML enhancements. This result corroborates with Hajra et al. (2023), reporting ~100 eV to 100s keV electron injections in the magnetotail during the SSS events. Based on the ESA/SaRIF reconstructed $L$*-shell variations of the ~MeV electrons, the FF shock impingement led to ~MeV electron flux losses in the outer radiation belt for $L$* = 3.5–5.5., and fluxes are enhanced in the lower $L$* region ($L$* ≤ 3.5). Hajra and Tsurutani (2018a) reported decreases in the geosynchronous orbit > 0.8 and > 2.0 MeV electron fluxes during the geomagnetic storm main phase, starting with the shock impingement. Compression of dayside magnetospheric magnetic fields by the FF shock makes them blunter than a normal dipole configuration. In fact, Dmitriev (2024) observed a long-lasting magnetosheath interval at the geosynchronous orbit, implying that the magnetopause was contracted below the geosynchronous orbit right after the FF shock arrival. As a result, energetic electrons gradient drifting from the midnight to dawn sector will drift towards the dayside magnetopause boundary and be lost to the magnetosheath. This process, called magnetopause shadowing, may lead to the electron losses on open drift paths (West et al., 1972, 1981; Li et al., 1997; Kim et al., 2008; Ohtani et al., 2009; Hietala et al., 2014; Hudson et al., 2014). Another possibility is the loss due to relativistic electron interaction with EMIC waves (Thorne & Kennel, 1971; Horne & Thorne, 1998; Summers et al., 1998; Meredith et al., 2006; Tsurutani et al., 2016). Solar wind ram pressure enhancement can excite EMIC waves in the dayside magnetosphere, and the cyclotron resonant interactions of the relativistic electrons with EMIC waves is shown to be a possible loss mechanism for these particles to the ionosphere (e.g., Remya et al., 2015; Tsurutani et al., 2016). Both these processes can be present during this case.

6.  The solar wind plasma rarefaction during the MC led to a sub-Alfvénic interval at the center of the MC. Chen et al. (2024) reported satellite observations of the Earth's magnetosphere interaction with this sub-Alfvénic solar wind. Based on a statistical survey, Hajra and Tsurutani (2022) suggested that Earth may encounter one sub-Alfvénic solar wind event every 2 years, and that the majority (83%) of the events are caused by superradial expansions of the outwardly propagating MCs, which is consistent with the present observation. The sub-Alfvénic solar wind did not cause any significant ring current variations. However, the sub-Alfvénic interval was associated with disappearance of auroral substorm occurrence, reduction in the EMIC wave activity, and reduction in ionospheric Joule heating rate. Why are EMIC waves reduced during the sub-Alfvénic solar wind interval? Due to the magnetospheric expansion, energetic ions in the magnetosphere will experience betatron de-acceleration. This will cause the ions to be stable against instability, and therefore no EMIC wave growth. In addition, there were no substorm onsets during the sub-Alfvénic solar wind impact. Disappearance of substorms indicates lack of fresh injections of anisotropic ~10–100 keV electrons and protons, which are important for wave generation. Thus, reduction in the EMIC wave activity during the sub-Alfvénic solar wind is due to both betatron de-acceleration of energetic magnetospheric protons and the absence of any new anisotropic source protons.

7.  During the geomagnetic storm main phase, the ionospheric plasma anomaly structure became more extended in latitude, and stronger in amplitude, with a ~60% increase in the plasma density and a ~80% increase in TEC, and ~33% and ~67% increases in their latitudinal extents, respectively, compared to their quiet-time values. The ionospheric plasma density and TEC variations are contributed by physical processes like plasma production, transport and loss (e.g., Hanson & Moffett, 1966). Plasma production from neutral particles through photo-ionization, and loss through chemical reactions among ions, electrons and neutral particles are dependent on ionospheric composition. The transport is controlled by equatorial electric field through the process of "fountain effect", that is, an upward drift of the plasma over magnetic equator followed by diffusion along the geomagnetic field lines on both sides of the magnetic equator (Martyn, 1955; Duncan, 1959). Inter-hemispheric plasma flow by atmospheric wind circulation may introduce significant hemispheric asymmetry in the plasma anomaly structure (Hanson & Moffett, 1966; Bailey et al., 1997). A detailed study on these various processes is beyond scope of this work. It has been shown that during a storm main phase, a PPEF reaches the equatorial *F*-region ionosphere during both daytime and nighttime (Tsurutani et al., 2004, 2007, 2008; Mannucci et al., 2005). On the other hand, this electric field is superposed on top of the diurnal electric field described above, and causes what is called the "dayside superfountain effect". The *F*-region ionospheric plasma can reach heights of ~1000 km or more (e.g., Tsurutani et al., 2004) and because of the $\boldsymbol{E} \times \boldsymbol{B}$ convection of the plasma, the ionospheric anomalies reach higher magnetic latitudes. During the 30–31 October 2003 Halloween superstorm, the anomalies reached ~±30° magnetic latitude (instead of the usual ~10° MLAT). The TEC within the ionospheric anomalies reached ~six times the normal values (Mannucci et al., 2005). What is the cause of the dayside ionospheric TEC enhancement? When the *F*-region plasma is lifted to higher altitudes, the recombination rate there is much lower than at lower altitudes. Therefore, the recombination of ions back into neutrals is substantially decreased. Meanwhile solar photons are creating new ionospheric plasma at lower altitudes replacing the plasma that has been uplifted, increasing TEC. The uplift of energetic $O^+$ ions can cause substantial satellite drag. For magnetic storms such as the Carrington magnetic storm, this could be catastrophic for low-altitude satellites (Tsurutani et al., 2012). There is an even worse scenario. For such extreme magnetic storm PPEFs, it is possible that ion-neutral drag will cause the uplift of substantial neutral O atoms as well (Lakhina & Tsurutani, 2017). We encourage modelers to attempt to evaluate how much satellite drag could occur during a Carrington-like event.

## 5 Concluding Remarks

The 23–24 April 2023 double-peak geomagnetic storm was associated with two strong IMF southward components provided by an interplanetary sheath followed by a MC, respectively. In addition, ion injection during several intense substorms and an SSS seem to largely contribute to the geomagnetic storm energetics. Equal amount of ionospheric Joule heating during dayside and nightside of the ionosphere advocates for a global-scale DP2 current system during the intense storm main phase. Loss of relativistic electrons from the outer radiation belt by the FF shock/sheath impact is consistent with earlier results. Enhancement of ionospheric plasma densities around the anomaly crests and latitudinal expansion of the equatorial ionospheric anomaly structures support the strengthening of equatorial fountain effect by prompt penetration of magnetospheric convection electric fields during storm main phase. This magnetic storm is certainly a classical one because it exhibits all of the major magnetospheric and ionospheric space weather effects known at this time. There may be many deleterious effects for humanity from this complex magnetic storm. We hope we have provided the plasma physics background for future reporting of these effects.

## Acknowledgments
The work of R. H. is funded by the "Hundred Talents Program" of the Chinese Academy of Sciences (CAS), and the Excellent Young Scientists Fund Program (Overseas) of the National Natural Science Foundation of China (NSFC). R. B. H. was supported by subcontract 68744/54.61/SWESNET-29-UKRI-BAS. Work at LPC2E and Lagrange is co-funded by CNES APR grants. We thank Dr. Claudia Stolle (IAP, Universität Rostock, Germany) for useful discussion on Swarm data. We thank Dr. Xavier Vallières (LPC2E, CNRS, France) for useful discussion on Cluster data. We thank the AMPERE team and the AMPERE Science Data Center for providing data products derived from the Iridium Communications constellation, enabled by support from the National Science Foundation. We thank the ESA Swarm Data Access for providing the Swarm data products. We would like to thank the reviewers for extremely valuable suggestions that substantially improved the manuscript.

## Open Research
### Data Availability Statement
Geomagnetic SYM-H indices are obtained from the World Data Center for Geomagnetism, Kyoto, Japan (http://wdc.kugi.kyoto-u.ac.jp/). The SYM-H indices can be freely downloaded from "Plot and data output of ASY/SYM and AE indices" (https://wdc.kugi.kyoto-u.ac.jp/aeasy/index.html) by specifying start time (UT) and duration. The SML indices are obtained from the SuperMAG website (https://supermag.jhuapl.edu/). The SML indices can be freely downloaded from the Indices (https://supermag.jhuapl.edu/indices/) by specifying the index, time, and duration. The solar wind plasma and IMF data are obtained from NASA's OMNIWeb (https://omniweb.gsfc.nasa.gov/). The CME information is obtained from the NASA's SOHO/LASCO CME catalog (https://cdaw.gsfc.nasa.gov/CME_list/halo/halo.html). The Cluster wave data are obtained from ESA/CSA (https://www.cosmos.esa.int/web/csa/access). Using the Cluster quicklook plots link (http://www.cluster.rl.ac.uk/csdsweb-cgi/csdsweb_pick) the plots can be viewed and downloaded by specifying the time period. The polar region FACs are obtained from AMPERE (https://ampere.jhuapl.edu/). The AMPERE Data Browsing and Plotting Tool (https://ampere.jhuapl.edu/browse/) can be used to view and download specific plots by specifying time period. The ~76 keV to ~2.9 MeV electron fluxes are obtained from the GOES-18 satellite (https://www.ngdc.noaa.gov/stp/satellite/goes-r.html). Specific GOES-18 data used in this work are freely available at the Directory: https://data.ngdc.noaa.gov/platforms/solar-space-observing-satellites/goes/goes18/l2/data/mpsh-l2-avg1m/2023/04/. The > 0.8 MeV and > 2.0 MeV electron $L^*$-shell variations are obtained from the ESA's SaRIF system (https://swe.ssa.esa.int/web/guest/sarif-federated). Researchers are required to create a free account to view and download SaRIF plots by specifying time through this site. The Swarm plasma density and TEC data are obtained from the Swarm Data Access (https://swarm-diss.eo.esa.int). Available Swaem Level 2 long-term data are used for this work.

**Author Contributions**
**Conceptualization**: Rajkumar Hajra, Bruce T. Tsurutani;
**Data curation**: Rajkumar Hajra, Xu Yang;
**Formal analysis**: Rajkumar Hajra;
**Funding acquisition**: Rajkumar Hajra;
**Investigation**: Rajkumar Hajra;
**Methodology**: Rajkumar Hajra;
**Resources**: Rajkumar Hajra;
**Supervision**: Bruce T. Tsurutani;
**Validation**: Rajkumar Hajra;
**Visualization**: Rajkumar Hajra, Xu Yang;
**Writing – original draft**: Rajkumar Hajra, Bruce T. Tsurutani;
**Writing – review & editing**: Rajkumar Hajra, Bruce T. Tsurutani, Quanming Lu, Richard B. Horne, Gurbax S. Lakhina, Xu Yang, Pierre Henri, Aimin Du, Xingliang Gao, Rongsheng Wang, San Lu

**Competing Interests**
The authors declare no competing interests.

**References**

1. Abraham-Shrauner, B. (1972). Determination of magnetohydrodynamic shock normal. *Journal of Geophysical Research, 77*, 736–739. https://doi.org/10.1029/JA077i004p00736
2. Akasofu, S. I. (1981). Energy coupling between the solar wind and the magnetosphere. *Space Science Reviews, 28*, 121–190. https://doi.org/10.1007/BF00218810
3. Anderson, B. J., Takahashi, K., Kamei, T., Waters, C. L., & Toth, B. A. (2002). Birkeland current system key parameters derived from Iridium observations: Method and initial validation results. *Journal of Geophysical Research, 107*, 1079. https://doi.org/10.1029/2001ja000080
4. Anderson, B. J., Angappan, R., Barik, A., Vines, S. K., Stanley, S., Bernasconi, P. N., Korth, H., & Barnes, R. J. (2021). Iridium communications satellite constellation data for study of Earth's magnetic field. *Geochemistry, Geophysics, Geosystems, 22*, e2020GC009515. https://doi.org/10.1029/2020GC009515

5. Appleton, E. V. (1946). Two anomalies in the ionosphere. *Nature, 157*, 691–693. https://doi.org/10.1038/157691a0

6. Axford, W. I., & Hines, C. O. (1961). A unifying theory of high-latitude geophysical phenomena and geomagnetic storms. *Canadian Journal of Physics, 39*, 1433–1464. https://doi.org/10.1139/p61-172

7. Bailey, G. J., Su, Y. Z., & Balan, N. (1997). The Sheffield University plasmasphere-ionosphere model–a review. *Journal of Atmospheric and Solar-Terrestrial Physics, 59*, 1541–1552. https://doi.org/10.1016/S1364-6826(96)00155-1

8. Baker, K., & Wing, S. (1989). A new magnetic coordinate system for conjugate studies at high latitudes. *Journal of Geophysical Research, 94*, 9139–9243. https://doi.org/10.1029/JA094iA07p09139

9. Baker, D. N., Blake, J. B., Callis, L. B., Cummings, J. R., Hovestadt, D., Kanekal, S., Klecker, B., Mewaldt, R. A., & Zwickl, R. D. (1994). Relativistic electron acceleration and decay time scales in the inner and outer radiation belts: SAMPEX. *Geophysical Research Letters, 21*, 409–412. https://doi.org/10.1029/93GL03532

10. Burlaga, L., Sittler, E., Mariani, F., & Schwenn, R. (1981). Magnetic loop behind an interplanetary shock: Voyager, Helios, and IMP 8 observations. *Journal of Geophysical Research, 86*, 6673–6684. https://doi.org/10.1029/JA086iA08p06673

11. Burton, R. K., McPherron, R. L., & Russell, C. T. (1975). An empirical relationship between interplanetary conditions and Dst. *Journal of Geophysical Research, 80*, 4204–4214. https://doi.org/10.1029/JA080i031p04204

12. Chané, E., Saur, J., Neubauer, F. M., Raeder, J., & Poedts, S. (2012). Observational evidence of Alfvén wings at the Earth. *Journal of Geophysical Research, 117*, A09217. https://doi.org/10.1029/2012JA017628

13. Chané, E., Raeder, J., Saur, J., Neubauer, F. M., Maynard, K. M., & Poedts, S. (2015). Simulations of the Earth's magnetosphere embedded in sub-Alfvénic solar wind on 24 and 25 May 2002. *Journal of Geophysical Research: Space Physics, 120*, 8517–8528. https://doi.org/10.1002/2015JA021515

14. Chapman, S., & Ferraro, V. C. A. (1931). A new theory of magnetic storms, part I, The initial phase. *Terrestrial Magnetism and Atmospheric Electricity, 36*, 77–97. https://doi.org/10.1029/TE036i002p00077

15. Chen, L.-J., Gershman, D., Burkholder, B., Chen, Y., Sarantos, M., Jian, L., et al. (2024). Earth's Alfvén wings driven by the April 2023 Coronal Mass Ejection. *Geophysical Research Letters, 51*, e2024GL108894. https://doi.org/10.1029/2024GL108894

16. Cornwall, J. M. (1965). Cyclotron instabilities and electromagnetic emission in the ultra low frequency and very low frequency ranges. *Journal of Geophysical Research, 70*, 61–69. https://doi.org/10.1029/JZ070i001p00061

17. Cummings, W. D., & Dessler, A. J. (1967). Field-aligned currents in the magnetosphere. *Journal of Geophysical Research, 72*, 1007–1013. https://doi.org/10.1029/JZ072i003p01007

18. Daglis, I. A. (1997). The role of magnetosphere-ionosphere coupling in magnetic storm dynamics. In: Magnetic Storms (eds B.T. Tsurutani, W.D. Gonzalez, Y. Kamide, & J.K. Arballo). https://doi.org/10.1029/GM098p0107

19. DeForest, S. E., & McIlwain, C. E. (1971). Plasma clouds in the magnetosphere. *Journal of Geophysical Research, 76*, 3587–3611. https://doi.org/10.1029/JA076i016p03587

20. Despirak, I., Setsko, P., Lubchich, A., Hajra, R., Sakharov, Y., Lakhina, G., Selivanov, V., & Tsurutani, B. T. (2024). Geomagnetically induced currents (GICs) during strong geomagnetic activity (storms, substorms, and magnetic pulsations) on 23–24 April 2023. *Journal of Atmospheric and Solar-Terrestrial Physics, 261*, 106293. https://doi.org/10.1016/j.jastp.2024.106293

21. Dessler, A. J., & Parker, E. N. (1959). Hydromagnetic theory of geomagnetic storms. *Journal of Geophysical Research, 64*, 2239–2252. https://doi.org/10.1029/JZ064i012p02239

22. Dmitriev, A. V. (2024). Geosynchronous Magnetopause Crossings in February–April 2023. *Cosmic Research, 62*, 220–230. https://doi.org/10.1134/S001095252360035X

23. Duncan, R. A. (1959). The equatorial F-region of the ionosphere. *Journal of Atmospheric and Terrestrial Physics, 18*, 89–100. https://doi.org/10.1016/0021-9169(60)90081-7

24. Dungey, J. W. (1961). Interplanetary magnetic field and the auroral zones, *Physical Review Letters, 6*, 47–48. https://doi.org/10.1103/PhysRevLett.6.47

25. Echer, E., Gonzalez, W. D., Tsurutani, B. T., & Gonzalez, A. L. C. (2008). Interplanetary conditions causing intense geomagnetic storms (Dst ≤ –100 nT) during solar cycle 23 (1996–2006). *Journal of Geophysical Research, 113*, A05221. https://doi.org/10.1029/2007JA012744

26. Fairfield, D. H., Iver, H. C., Desch, M. D., Szabo, A., Lazarus, A. J., & Aellig, M. R. (2001). The location of low Mach number bow shocks at Earth. *Journal of Geophysical Research, 106*, 25361–25376. https://doi.org/10.1029/2000JA000252

27. Farris, M. H., & Russell, C. T. (1994). Determining the standoff distance of the bow shock: Mach number dependence and use of models. *Journal of Geophysical Research, 99*, 17681–17689. https://doi.org/10.1029/94JA01020

28. Gjerloev, J. W. (2009). A global ground-based magnetometer initiative. *Eos Transactions American Geophysical Union, 90*, 230–231. https://doi.org/10.1029/2009EO270002

29. Gjerloev, J. W. (2012). The SuperMAG data processing technique. *Journal of Geophysical Research, 117*, A09213. https://doi.org/10.1029/2012JA017683

30. Glauert, S. A., Horne, R. B., & Meredith, N. P. (2014). Three dimensional electron radiation belt simulations using the BAS Radiation Belt Model with new diffusion models for chorus, plasmaspheric hiss and lightning-generated whistlers. *Journal of Geophysical Research: Space Physics, 119*, 268–289. https://doi.org/10.1002/2013JA019281

31. Gonzalez, W. D., & Tsurutani, B. T. (1987). Criteria of interplanetary parameters causing intense magnetic storms (Dst < −100 nT). *Planetary & Space Science, 35*, 1101–1109. https://doi.org/10.1016/0032-0633(87)90015-8

32. Gonzalez, W. D., Tsurutani, B. T., Gonzalez, A. L. C., Smith, E. J., Tang, F., & Akasofu, S. I. (1989). Solar wind-magnetosphere coupling during intense magnetic storms (1978-1979). *Journal of Geophysical Research, 94*, 8835–8851. https://doi.org/10.1029/JA094iA07p08835

33. Gonzalez, W. D., Joselyn, J. A., Kamide, Y., Kroehl, H. W., Rostoker, G., Tsurutani, B. T., & Vasyliunas, V. M. (1994). What is a geomagnetic storm? *Journal of Geophysical Research, 99*, 5771–5792. https://doi.org/10.1029/93JA02867

34. Gonzalez, W. D., Gonzalez, A. L. C., Dal Lago, A., Tsurutani, B. T., Arballo, J. K., Lakhina, G. S., Buti, B., Ho, C. M., & Wu, S.-T. (1998). Magnetic cloud field intensities and solar wind velocities. *Geophysical Research Letters, 25*, 963–966. https://doi.org/10.1029/98GL00703

35. Gosling, J. T., Asbridge, J. R., Bame, S. J., Feldman, W. C., Zwickl, R. D., Paschmann, G., Sckopke, N., & Russell, C. T. (1982). A Sub-Alfvénic solar wind: Interplanetary and magnetosheath observations. *Journal of Geophysical Research, 87*, 239–245. https://doi.org/10.1029/JA087iA01p00239

36.     Guo, J., Feng, X., Emery, B. A., Zhang, J., Xiang, C., Shen, F., & Song, W. (2011). Energy transfer during intense geomagnetic storms driven by interplanetary coronal mass ejections and their sheath regions. *Journal of Geophysical Research, 116*, A05106. https://doi.org/10.1029/2011JA016490

37.     Guo, J., Feng, X., Emery, B. A., & Wang, Y. (2012). Efficiency of solar wind energy coupling to the ionosphere. *Journal of Geophysical Research, 117*, A07303. https://doi.org/10.1029/2012JA017627

38.     Hajra, R., Echer, E., Tsurutani, B. T., & Gonzalez, W. D. (2013). Solar cycle dependence of High-Intensity Long-Duration Continuous AE Activity (HILDCAA) events, relativistic electron predictors? *Journal of Geophysical Research: Space Physics, 118*, 5626–5638. https://doi.org/10.1002/jgra.50530

39.     Hajra, R., Echer, E., Tsurutani, B. T., & Gonzalez, W. D. (2014). Solar wind-magnetosphere energy coupling efficiency and partitioning: HILDCAAs and preceding CIR storms during solar cycle 23. *Journal of Geophysical Research: Space Physics, 119*, 2675–2690. https://doi.org/10.1002/2013JA019646

40.     Hajra, R., Tsurutani, B. T., Echer, E., Gonzalez, W. D., & Gjerloev, J. W. (2016). Supersubstorms (SML < –2500 nT): Magnetic storm and solar cycle dependences. *Journal of Geophysical Research: Space Physics, 121*, 7805–7816. https://doi.org/10.1002/2015JA021835

41.     Hajra, R., Henri, P., Vallières, X., Galand, M., Héritier, K., Eriksson, A. I., Odelstad, E., Edberg, N. J. T., Burch, J. L., Broiles, T., Goldstein, R., Glassmeier, K. H., Richter, I., Goetz, C., Tsurutani, B. T., Nilsson, H., Altwegg, K., & Rubin, M. (2017). Impact of a cometary outburst on its ionosphere: Rosetta Plasma Consortium observations of the outburst exhibited by comet 67P/Churyumov-Gerasimenko on 19 February 2016. *Astronomy & Astrophysics, 607*, A34. https://doi.org/10.1051/0004-6361/201730591

42.     Hajra, R., & Tsurutani, B. T. (2018a). Chapter 14 – Magnetospheric "killer" relativistic electron dropouts (REDs) and repopulation: a cyclical process. In: Extreme Events in Geospace (eds. N. Buzulukova), pages 373–400, Elsevier. https://doi.org/10.1016/B978-0-12-812700-1.00014-5

43.     Hajra, R., & Tsurutani, B. T. (2018b). Interplanetary Shocks Inducing Magnetospheric Supersubstorms (SML < −2500 nT): Unusual Auroral Morphologies and Energy Flow. *The Astrophysical Journal, 858*, 123. https://doi.org/10.3847/1538-4357/aabaed

44.     Hajra, R., Tsurutani, B. T., & Lakhina, G. S. (2020). The complex space weather events of 2017 September. *The Astrophysical Journal, 899*, 3. https://doi.org/10.3847/1538-4357/aba2c5

45.     Hajra, R. (2021). Variation of the interplanetary shocks in the inner heliosphere. *The Astrophysical Journal, 917*, 91. https://doi.org/10.3847/1538-4357/ac0897

46.     Hajra, R., Sunny, J. V., Babu, M., Nair, A. G. (2022). Interplanetary Sheaths and Corotating Interaction Regions: A Comparative Statistical Study on Their Characteristics and Geoeffectiveness. *Solar Physics, 297*, 97. https://doi.org/10.1007/s11207-022-02020-6

47.     Hajra, R., & Tsurutani, B. T. (2022). Near-Earth sub-Alfvénic solar winds: interplanetary origins and geomagnetic impacts. *The Astrophysical Journal, 926*, 135. https://doi.org/10.3847/1538-4357/ac4471

48.     Hajra, R., Echer, E., Franco, A. M. S., & Bolzan, M. J. A. (2023). Earth's magnetotail variability during supersubstorms (SSSs): A study on solar wind–magnetosphere–ionosphere coupling. *Advances in Space Research, 72*, 1208–1223. https://doi.org/10.1016/j.asr.2023.04.013

49.     Hajra, R., Tsurutani, B. T., Lakhina, G. S., Lu, Q., & Du, A. (2024). Interplanetary Causes and Impacts of the 2024 May Superstorm on the Geosphere: An Overview. *The Astrophysical Journal*. https://doi.org/10.3847/1538-4357/ad7462

50.     Hanson, W. B., & Moffett, R. J. (1966). Ionization transport effects in the equatorial *F* region. *Journal of Geophysical Research, 71*, 5559–5572. https://doi.org/10.1029/JZ071i023p05559

51.     Hietala, H., Kilpua, E. K. J., Turner, D. L., & Angelopoulos, V. (2014). Depleting effects of ICME-driven sheath regions on the outer electron radiation belt. *Geophysical Research Letters, 41*, 2258–2265. https://doi.org/10.1002/2014GL059551

52.     Horne, R. B., & Thorne, R. M. (1998). Potential waves for relativistic electron scattering and stochastic acceleration during magnetic storms. *Geophysical Research Letters, 25*, 3011–3014. https://doi.org/10.1029/98GL01002

53.     Horne, R. B., Thorne, R. M., Shprits, Y. Y., Meredith, N. P., Glauert, S. A., Smith, A. J., Kanekal, S. G., Baker, D. N., Engebretson, M. J., Posch, J. L., Spasojevic, M., Inan, U. S., Pickett, J. S., Decreau, P. M. E. (2005). Wave acceleration of electrons in the Van Allen radiation belts. *Nature, 437*, 227–230. https://doi.org/10.1038/nature03939

54.     Horne, R. B., Lam, M. M., & Green, J. C. (2009). Energetic electron precipitation from the outer radiation belt during geomagnetic storms. *Geophysical Research Letters, 36*, L19104. https://doi.org/10.1029/2009GL040236

55.     Horne, R. B., Glauert, S. A., Meredith, N. P., Heynderickx, D., Boscher, D., Maget, V., & Pitchford, D. (2013). Space weather impacts on satellites and forecasting the Earth's electron radiation belts with SPACECAST. *Space Weather, 11*, 169–186. https://doi.org/10.1002/swe.20023

56.     Horne, R. B., Glauert, S. A., Kirsch, P., Heynderickx, D., Bingham, S., Thorn, P., Curran, B.-A., Pitchford, D., Haggarty, E., Wade, D., & Keil, R. (2021). The satellite risk prediction and radiation forecast system (SaRIF). *Space Weather, 19*, e2021SW002823. https://doi.org/10.1029/2021SW002823

57.     Hudson, M. K., Baker, D. N., Goldstein, J., Kress, B. T., Paral, J., Toffoletto, F. R., & Wiltberger, M. (2014). Simulated magnetopause losses and Van Allen Probe flux dropouts. *Geophysical Research Letters, 41*, 1113–1118. https://doi.org/10.1002/2014GL059222

58.     Hugoniot, H. (1887). Mémoire sur la propagation des mouvements dans les corps et spécialement dans les gaz parfaits (première partie). *Journal de l'École Polytechnique, 57*, 3–97.

59.     Hugoniot, H. (1889). Mémoire sur la propagation des mouvements dans les corps et spécialement dans les gaz parfaits (deuxième partie). *Journal de l'École Polytechnique, 58*, 1–125.

60.     Iyemori, T. (1990). Storm-time magnetospheric currents inferred from mid-latitutde geomagnetic field variations. *Journal of Geomagnetism and Geoelectricity, 42*, 1249–1265. https://doi.org/10.5636/jgg.42.1249

61.     Kamide, Y., Yokoyama, N., Gonzalez, W., Tsurutani, B. T., Daglis, I. A., Brekke, A., & Masuda, S. (1998). Two-step development of geomagnetic storms. *Journal of Geophysical Research, 103*, 6917–6921. https://doi.org/10.1029/97JA03337

62.      Kang, N., Lu, Q., Gao, X., Wang, X., Chen, H., & Wang, S. (2021). Propagation of electromagnetic ion cyclotron waves in a dipole magnetic field: A 2-D hybrid simulation. *Journal of Geophysical Research: Space Physics, 126*, e2021JA029720. https://doi.org/10.1029/2021JA029720

63.      Kennel, C. F., & Petschek, H. E. (1966). Limit on stably trapped particle fluxes. *Journal of Geophysical Research, 71*, 1–28. https://doi.org/10.1029/JZ071i001p00001

64.      Kennel, C. F., Edmiston, J. P., & Hada, T. (1985). A Quarter Century of Collisionless Shock Research. In: Collisionless Shocks in the Heliosphere: A Tutorial Review (eds R. G. Stone, & B. T. Tsurutani). https://doi.org/10.1029/GM034p0001

65.      Kim, K. C., Lee, D.-Y., Kim, H.-J., Lyons, L. R., Lee, E. S., Öztürk, M. K., & Choi, C. R. (2008). Numerical calculations of relativistic electron drift loss effect. *Journal of Geophysical Research, 113*, A09212. https://doi.org/10.1029/2007JA013011

66.      Knipp, D. J., Tobiska, W. K., & Emery, B. A. (2004). Direct and indirect thermospheric heating sources for solar cycles 21–23. *Solar Physics, 224*, 495–504. https://doi.org/10.1007/s11207-005-6393-4

67.      Knudsen, D. J., Burchill, J. K., Buchert, S. C., Eriksson, A. I., Gill, R., Wahlund, J.-E., Åhlen, L., Smith, M., & Moffat, B. (2017). Thermal ion imagers and Langmuir probes in the Swarm electric field instruments. *Journal of Geophysical Research: Space Physics, 122*, 2655–2673. https://doi.org/10.1002/2016JA022571

68.      Lakhina, G. S., & Tsurutani, B. T. (2017). Satellite drag effects due to uplifted oxygen neutrals during super magnetic storms. *Nonlinear Processes in Geophysics, 24*, 745–750. https://doi.org/10.5194/npg-24-745-2017

69.      Li, X., Baker, D. N., Temerin, M., Cayton, T. E., Reeves, E. G. D., Christensen, R. A., Blake, J. B., Looper, M. D., Nakamura, R., & Kanekal, S. G. (1997). Multisatellite observations of the outer zone electron variation during the November 3–4, 1993, magnetic storm. *Journal of Geophysical Research, 102*, 14123–14140. https://doi.org/10.1029/97JA01101

70.      Lu, Q. M., Guo, F., & Wang, S. (2006). Magnetic spectral signatures in the terrestrial plasma depletion layer: Hybrid simulations. *Journal of Geophysical Research, 111*, A04207. https://doi.org/10.1029/2005JA011405

71.      Lu, Q., Fu, H., Wang, R., & Lu, S. (2022). Collisionless magnetic reconnection in the magnetosphere. *Chinese Physics B*, 31, 089401. https://doi.org/10.1088/1674-1056/ac76ab

72.      Lugaz, N., Farrugia, C. J., Huang, C.-L., Winslow, R. M., Spence, H. E., & Schwadron, N. A. (2016). Earth's magnetosphere and outer radiation belt under sub-Alfvénic solar wind. *Nature Communications, 7*, 13001. https://doi.org/10.1038/ncomms13001

73.      Mannucci, A. J., Tsurutani, B. T., Iijima, B. A., Komjathy, A., Saito, A., Gonzalez, W. D., Guarnieri, F. L., Kozyra, J. U., & Skoug, R. (2005). Dayside global ionospheric response to the major interplanetary events of October 29–30, 2003 "Halloween storms". *Geophysical Research Letters, 32*, L12S02. https://doi.org/10.1029/2004GL021467

74.      Martyn, D. F. (1955). Geomagnetic anomalies of the F2 region and their interpretation, *The Physics of the Ionosphere*, Physical Society, London. 260–264.

75.      Marubashi, K., & Lepping, R. P. (2007). Long-duration magnetic clouds: a comparison of analyses using torus- and cylinder-shaped flux rope models. Annales Geophysicae, 25, 2453 – 2477. https://doi.org/10.5194/angeo-25-2453-2007

76.     McIlwain, C. E. (1961). Coordinates for mapping the distribution of magnetically trapped particles. *Journal of Geophysical Research, 66*, 3681–3691. https://doi.org/10.1029/JZ066i011p03681

77.     McPherron, R. L. (1997). The role of substorms in the generation of magnetic storms. In: Magnetic Storms (eds B.T. Tsurutani, W.D. Gonzalez, Y. Kamide, & J.K. Arballo). https://doi.org/10.1029/GM098p0131

78.     Meng, X., Tsurutani, B. T., & Mannucci, A. J. (2019). The solar and interplanetary causes of superstorms (minimum Dst ≤ −250 nT) during the space age. *Journal of Geophysical Research: Space Physics, 124*, 3926–3948. https://doi.org/10.1029/2018JA026425

79.     Meredith, N. P., Horne, R. B., Glauert, S. A., Thorne, R. M., Summers, D., Albert, J. M., & Anderson, R. R. (2006). Energetic outer zone electron loss timescales during low geomagnetic activity. *Journal of Geophysical Research, 111*, A05212. https://doi.org/10.1029/2005JA011516

80.     Monreal-MacMahon, R., & Gonzalez, W. D. (1997). Energetics during the main phase of geomagnetic superstorms. *Journal of Geophysical Research, 102*, 14199–14207. https://doi.org/10.1029/97JA01151

81.     Newell, P. T., & Gjerloev, J. W. (2011a). Evaluation of SuperMAG auroral electrojet indices as indicators of substorms and auroral power. *Journal of Geophysical Research, 116*, A12211. https://doi.org/10.1029/2011JA016779

82.     Newell, P. T., & Gjerloev, J. W. (2011b). Substorm and magnetosphere characteristic scales inferred from the SuperMAG auroral electrojet indices. *Journal of Geophysical Research, 116*, A12232. https://doi.org/10.1029/2011JA016936

83.     Nishida, A. (1968). Geomagnetic Dp 2 fluctuations and associated magnetospheric phenomena. *Journal of Geophysical Research, 73*, 1795–1803. https://doi.org/10.1029/JA073i005p01795

84.     Ohtani, S., Miyoshi, Y., Singer, H. J., & Weygand, J. M. (2009). On the loss of relativistic electrons at geosynchronous altitude: Its dependence on magnetic configurations and external conditions. *Journal of Geophysical Research, 114*, A01202. https://doi.org/10.1029/2008JA013391

85.     Olsen, N., Friis-Christensen, E., Floberghagen, R., Alken, P., Beggan, C. D., Chulliat, A., Doornbos, E., Encarnação, J. T., Hamilton, B., Hulot, G., van den Ijssel, J., Kuvshinov, A., Lesur, V., Lühr, H., Macmillan, S., Maus, S., Noja, M., Olsen, P. E. H., Park, J., Plank, G., Püthe, C., Rauberg, J., Ritter, P., Rother, M., Sabaka, T. J., Schachtschneider, R., Sirol, O., Stolle, C., Thébault, E., Thomson, A. W. P., Tøffner-Clausen, L., Velímský, J., Vigneron, P., & Visser, P. N. (2013). The swarm satellite constellation application and research facility (SCARF) and swarm data products. *Earth Planets and Space, 65*, 1189–1200. https://doi.org/10.5047/eps.2013.07.001

86.     Onsager, T. G., Rostoker, G., Kim, H.-J., Reeves, G. D., Obara, T., Singer, H. J., & Smithtro, C. (2002). Radiation belt electron flux dropouts: Local time, radial, and particle-energy dependence. *Journal of Geophysical Research, 107*, 1382. https://doi.org/10.1029/2001JA000187

87.     Østgaard, N., Germany, G., Stadsnes, J., & Vondrak, R. R. (2002). Energy analysis of substorms based on remote sensing techniques, solar wind measurements, and geomagnetic indices. *Journal of Geophysical Research, 107*, 1233. https://doi.org/10.1029/2001JA002002

88.    Perreault, P., & Akasofu, S. I. (1978). A study of geomagnetic storms. *Geophysical Journal of the Royal Astronomical Society, 54*, 547–583. https://doi.org/10.1111/j.1365-246X.1978.tb05494.x

89.    Rankine, W. J. M. (1870). XV. On the thermodynamic theory of waves of finite longitudinal disturbance. *Philosophical Transactions of the Royal Society, 160*, 277–288. http://doi.org/10.1098/rstl.1870.0015

90.    Reeves, G. D., McAdams, K. L., Friedel, R. H. W., & O'Brien, T. P. (2003). Acceleration and loss of relativistic electrons during geomagnetic storms. *Geophysical Research Letters, 30*, 1529. https://doi.org/10.1029/2002GL016513

91.    Remya, B., Tsurutani, B. T., Reddy, R. V., Lakhina, G. S., & Hajra, R. (2015). Electromagnetic cyclotron waves in the dayside subsolar outer magnetosphere generated by enhanced solar wind pressure: EMIC wave coherency. *Journal of Geophysical Research: Space Physics, 120*, 7536–7551. https://doi.org/10.1002/2015JA021327

92.    Richmond, A. D., & Lu, G. (2000). Upper-atmospheric effects of magnetic storms: a brief tutorial. *Journal of Atmospheric and Solar-Terrestrial Physics, 62*, 1115 – 1127. https://doi.org/10.1016/S1364-6826(00)00094-8

93.    Roederer, J. G. (1970). Dynamics of Geomagnetically Trapped Radiation. Vol. 2, Springer, Berlin, Heidelberg. https://doi.org/10.1007/978-3-642-49300-3

94.    Sckopke, N. (1966). A general relation between the energy of trapped particles and the disturbance field near the Earth. *Journal of Geophysical Research, 71*, 3125–3130. https://doi.org/10.1029/JZ071i013p03125

95.    Shue, J.-H., Song, P., Russell, C. T., Steinberg, J. T., Chao, J. K., Zastenker, G., Vaisberg, O. L., Kokubun, S., Singer, H. J., Detman, T. R., & Kawano, H. (1998). Magnetopause location under extreme solar wind conditions. *Journal of Geophysical Research, 103*, 17691–17700. https://doi.org/10.1029/98JA01103

96.    Shue, J. H., & Chao, J. K. (2013). The role of enhanced thermal pressure in the earthward motion of the Earth's magnetopause. *Journal of Geophysical Research: Space Physics, 118*, 3017–3026. https://doi.org/10.1002/jgra.50290

97.    Smith, E. J. (1985). Interplanetary Shock Phenomena Beyond 1 AU. In Collisionless Shocks in the Heliosphere: Reviews of Current Research (eds B.T. Tsurutani and R.G. Stone). https://doi.org/10.1029/GM035p0069

98.    Smith, C. W., Mullan, D. J., Ness, N. F., Skoug, R. M., & Steinberg, J. (2001). Day the solar wind almost disappeared: Magnetic field fluctuations, wave refraction and dissipation. *Journal of Geophysical Research, 106*, 18625–18634. https://doi.org/10.1029/2001JA000022

99.    Summers, D., Thorne, R. M., & Xiao, F. (1998). Relativistic theory of wave-particle resonant diffusion with application to electron acceleration in the magnetosphere. *Journal of Geophysical Research, 103*, 20487–20500. https://doi.org/10.1029/98JA01740

100.   Tanaka, T., & Hirao, K. (1973). Effects of an electric field on the dynamical behavior of the ionospheres and its application to the storm time disturbance of the F-layer. *Journal of Atmospheric and Terrestrial Physics, 35*, 1443–1452. https://doi.org/10.1016/0021-9169(73)90147-5

101.   Tanskanen, E., Pulkkinen, T. I., Koskinen, H. E. J., & Slavin, J. A. (2002). Substorm energy budget during low and high solar activity: 1997 and 1999 compared. *Journal of Geophysical Research, 107*, SMP 15-1–SMP 15-11. https://doi.org/10.1029/2001JA900153

102. Tenfjord, P., & Østgaard, N. (2013). Energy transfer and flow in the solar wind-magnetosphere-ionosphere system: A new coupling function. *Journal of Geophysical Research: Space Physics, 118*, 5659–5672. https://doi.org/10.1002/jgra.50545

103. Thorne, R. M., & Kennel, C. F. (1971). Relativistic electron precipitation during magnetic storm main phase. *Journal of Geophysical Research, 76*, 4446–4453. https://doi.org/10.1029/JA076i019p04446

104. Tsurutani, B. T., & Meng, C.-I. (1972). Interplanetary magnetic-field variations and substorm activity. *Journal of Geophysical Research, 77*, 2964–2970. https://doi.org/10.1029/JA077i016p02964

105. Tsurutani, B. T., Smith, E. J., Anderson, R. R., Ogilvie, K. W., Scudder, J. D., Baker, D. N., & Bame, S. J. (1982). Lion roars and nonoscillatory drift mirror waves in the magnetosheath. *Journal of Geophysical Research, 87*, 6060–6072. https://doi.org/10.1029/JA087iA08p06060

106. Tsurutani, B. T., & Lin, R. P. (1985). Acceleration of >47 keV Ions and >2 keV electrons by interplanetary shocks at 1 AU. *Journal of Geophysical Research, 90*, 1–11. https://doi.org/10.1029/JA090iA01p00001

107. Tsurutani, B. T., Gonzalez, W. D., Tang, F., Akasofu, S. I., & Smith, E. J. (1988). Origin of interplanetary southward magnetic fields responsible for major magnetic storms near solar maximum (1978–1979). *Journal of Geophysical Research, 93*, 8519–8531. https://doi.org/10.1029/JA093iA08p08519

108. Tsurutani, B. T., Dasgupta, B., Arballo, J. K., Lakhina, G. S., & Pickett, J. S. (2003). Magnetic field turbulence, electron heating, magnetic holes, proton cyclotron waves, and the onsets of bipolar pulse (electron hole) events: A possible unifying scenario. *Nonlinear Processes in Geophysics, 21*, 27–35. https://doi.org/10.5194/npg-10-27-2003

109. Tsurutani, B. T., Mannucci, A., Iijima, B., Abdu, M. A., Sobral, J. H. A., Gonzalez, W., Guarnieri, F., Tsuda, T., Saito, A., Yumoto, K., Fejer, B., Fuller-Rowell, T. J., Kozyra, J., Foster, J. C., Coster, A., Vasyliunas, V. M. (2004). Global dayside ionospheric uplift and enhancement associated with interplanetary electric fields. *Journal of Geophysical Research, 109*, A08302. https://doi.org/10.1029/2003JA010342

110. Tsurutani, B. T., Verkhoglyadova, O. P., Mannucci, A. J., Araki, T., Saito, A., Tsuda, T., & Yumoto, K. (2007). Oxygen ion uplift and satellite drag effects during the 30 October 2003 daytime superfountain event. *Annales Geophysicae, 25*, 569–574. https://doi.org/10.5194/angeo-25-569-2007

111. Tsurutani, B. T., Verkhoglyadova, O. P., Mannucci, A. J., Saito, A., Araki, T., Yumoto, K., Tsuda, T., Abdu, M. A., Sobral, J. H. A., Gonzalez, W. D., McCreadie, H., Lakhina, G. S., & Vasyliūnas, V. M. (2008). Prompt penetration electric fields (PPEFs) and their ionospheric effects during the great magnetic storm of 30–31 October 2003. *Journal of Geophysical Research, 113*, A05311. https://doi.org/10.1029/2007JA012879

112. Tsurutani, B. T., Lakhina, G. S., Verkhoglyadova, O. P., Gonzalez, W. D., Echer, E., & Guarnieri, F. L. (2011a). A review of interplanetary discontinuities and their geomagnetic effects. *Journal of Atmospheric and Solar-Terrestrial Physics, 73*, 5–19. https://doi.org/10.1016/j.jastp.2010.04.001

113. Tsurutani, B. T., Echer, E., Guarnieri, F. L., & Gonzalez, W. D. (2011b). The properties of two solar wind high speed streams and related geomagnetic activity during the declining phase of solar cycle 23. *Journal of Atmospheric and Solar-Terrestrial Physics, 73*, 164–177. https://doi.org/10.1016/j.jastp.2010.04.003

114.    Tsurutani, B. T., Verkhoglyadova, O.P., Mannucci, A. J., Lakhina, G. S., & Huba, J. D. (2012). Extreme changes in the dayside ionosphere during a Carrington-type magnetic storm. *Journal of Space Weather and Space Climate, 2*, A05. https://doi.org/10.1051/swsc/2012004

115.    Tsurutani, B. T., Hajra, R., Echer, E., & Gjerloev, J. W. (2015). Extremely intense (SML ≤−2500 nT) substorms: isolated events that are externally triggered? *Annales Geophysicae, 33*, 519–524. https://doi.org/10.5194/angeo-33-519-2015

116.    Tsurutani, B. T., Hajra, R., Tanimori, T., Takada, A., Remya, B., Mannucci, A. J., Lakhina, G. S., Kozyra, J. U., Shiokawa, K., Lee, L. C., Echer, E., Reddy, R. V., & Gonzalez. W. D. (2016). Heliospheric plasma sheet (HPS) impingement onto the magnetosphere as a cause of relativistic electron dropouts (REDs) via coherent EMIC wave scattering with possible consequences for climate change mechanisms. *Journal of Geophysical Research: Space Physics, 121*, 10130–10156. https://doi.org/10.1002/2016JA022499

117.    Tsurutani, B. T., Lakhina, G. S., & Hajra, R. (2020). The physics of space weather/solar-terrestrial physics (STP): what we know now and what the current and future challenges are. *Nonlinear Processes in Geophysics, 27*, 75–119. https://doi.org/10.5194/npg-27-75-2020

118.    Tsurutani, B. T., Green, J., & Hajra, R. (2022). The Possible Cause of the 40 SpaceX Starlink Satellite Losses in February 2022: Prompt Penetrating Electric Fields and the Dayside Equatorial and Midlatitude Ionospheric Convective Uplift. arXiv preprint. https://arxiv.org/abs/2210.07902

119.    Tsurutani, B. T., & Hajra, R. (2023). Energetics of Shock-triggered Supersubstorms (SML<−2500 nT). *The Astrophysical Journal, 946*, 17. https://doi.org/10.3847/1538-4357/acb143

120.    Tsurutani, B. T., Zank, G. P., Sterken, V. J., Shibata, K., Nagai, T., Mannucci, A.J., Malaspina, D. M., Lakhina, G. S., Kanekal, S. G., Hosokawa, K., Horne, R. B., Hajra, R., Glassmeier, K.-H., Gaunt, C. T., Chen, P.-F., & Akasofu, S.-I. (2023). Space Plasma Physics: A Review. *IEEE Transactions on Plasma Science, 51*, 1595–1655. https://doi.org/10.1109/TPS.2022.3208906

121.    Tsurutani, B. T., Sen, A., Hajra, R., Lakhina, G. S., Horne, R. B., & Hada, T. (2024). Review of the August 1972 and March 1989 (Allen) space weather events: Can we learn anything new from them? *Journal of Geophysical Research: Space Physics, 129*, e2024JA032622. https://doi.org/10.1029/2024JA032622

122.    Turner, N. E., Baker, D. N., Pulkkinen, T. I., Roeder, J. L., Fennell, J. F., & Jordanova, V. K. (2001). Energy content in the storm time ring current. *Journal of Geophysical Research, 106*, 19149–19156. https://doi.org/10.1029/2000JA003025

123.    Turner, N. E., Mitchell, E. J., Knipp, D. J., & Emery, B. A. (2006). Energetics of magnetic storms driven by corotating interaction regions: A study of geoeffectiveness. In: Recurrent Magnetic Storms: Corotating Solar Wind Streams, Geophysical Monograph Series, vol. 167, edited by B. T. Tsurutani et al., 113 pp., AGU, Washington, D. C., https://doi.org/10.1029/167GM11

124.    Turner, N. E., Cramer, W. D., Earles, S. K., & Emery, B. A. (2009). Geoefficiency and energy partitioning in CIR-driven and CME-driven storms. *Journal of Atmospheric and Solar-Terrestrial Physics, 71*, 1023–1031. https://doi.org/10.1016/j.jastp.2009.02.005

125.    Turner, D. L., Angelopoulos, V., Li, W., Hartinger, M. D., Usanova, M., Mann, I. R., Bortnik, J., & Shprits, Y. (2013). On the storm-time evolution of relativistic electron phase space density in Earth's outer radiation belt. *Journal of Geophysical Research, 118*, 2196–2212. https://doi.org/10.1002/jgra.50151

126.   Turner, D. L., Angelopoulos, V., Li, W., Bortnik, J., Ni, B., Ma, Q., Thorne, R. M., Morley, S. K., Henderson, M. G., Reeves, G. D., Usanova, M., Mann, I. R., Claudepierre, S. G., Blake, J. B., Baker, D. N., Huang, C. L., Spence, H., Kurth, W., Kletzing, C., & Rodriguez, J. V. (2014). Competing source and loss mechanisms due to wave-particle interactions in Earth's outer radiation belt during the 30 September to 3 October 2012 geomagnetic storm. *Journal of Geophysical Research, 119*, 1960–1979. https://doi.org/10.1002/2014JA019770

127.   Usmanov, A. V., Goldstein, M. L., & Farrell, W. M. (2000). A view of the inner heliosphere during the May 10–11, 1999 low density anomaly. *Geophysical Research Letters, 27*, 3765–3768. https://doi.org/10.1029/2000GL000082

128.   Usmanov, A. V., Goldstein, M. L., Ogilvie, K. W., Farrell, W. M., & Lawrence, G. R. (2005). Low-density anomalies and sub-Alfvénic solar wind. *Journal of Geophysical Research, 110*, A01106. https://doi.org/10.1029/2004JA010699

129.   Vichare, G., Alex, S., & Lakhina, G. S. (2005). Some characteristics of intense geomagnetic storms and their energy budget. *Journal of Geophysical Research, 110*, A03204. https://doi.org/10.1029/2004JA010418

130.   von Humboldt, A. (1808). Die vollständigste aller bisherigen Beobachtungen über den Einfluss des Nordlichts auf die Magnetnadel angestellt. *Annalen der Physik, 29*, 425–429. https://doi.org/10.1002/andp.18080290806

131.   Waters, C. L., Anderson, B. J., & Liou, K. (2001). Estimation of global field aligned currents using the Iridium® system magnetometer data. *Geophysical Research Letters, 28*, 2165–2168. https://doi.org/10.1029/2000GL012725

132.   Waters, C. L., Anderson, B. J., Green, D. L., Korth, H., Barnes, R. J., & Vanhamäki, H. (2020). Chapter 7 – Science Data Products for AMPERE. In: Ionospheric Multi-Spacecraft Analysis Tools (eds. M. W. Dunlop, & H. Lühr), ISSI Scientific Report Series 17. https://doi.org/10.1007/978-3-030-26732-2_7

133.   West, H. I., Buck, R. M., & Walton, J. R. (1972). Shadowing of electron azimuthal-drift motions near the noon magnetopause. *Nature Physical Science 240*, 6–7. https://doi.org/10.1038/physci240006a0

134.   West Jr., H. I., Buck, R. M., & Davidson, G. T. (1981). The dynamics of energetic electrons in the Earth's outer radiation belt during 1968 as observed by the Lawrence Livermore National Laboratory's Spectrometer on Ogo 5. *Journal of Geophysical Research, 86*, 2111–2142. https://doi.org/10.1029/JA086iA04p02111

135.   Wygant, J. R., Bonnell, J. W., Goetz, K., Ergun, R. E., Mozer, F. S., Bale, S. D., Ludlam, M., Turin, P., Harvey, P. R., Hochmann, R., Harps, K., Dalton, G., McCauley, J., Rachelson, W., Gordon, D., Donakowski, B., Shultz, C., Smith, C., Diaz-Aguado, M., Fischer, J., Heavner, S., Berg, P., Malsapina, D. M., Bolton, M. K., Hudson, M., Strangeway, R. J., Baker, D. N., Li, X., Albert, J., Foster, J. C., Chaston, C. C., Mann, I., Donovan, E., Cully, C. M., Cattell, C. A., Krasnoselskikh, V., Kersten, K., Brenneman, A., & Tao, J. B. (2013). The electric field and waves instruments on the radiation belt storm probes mission. *Space Science Reviews, 179*, 183–220. https://doi.org/10.1007/s11214-013-0013-7

136.   Yokoyama, N., & Kamide, Y. (1997). Statistical nature of geomagnetic storms. *Journal of Geophysical Research, 102*, 14215–14222. https://doi.org/10.1029/97JA00903

137.   Yue, X., Schreiner, W. S., Lin, Y.-C., Rocken, C., Kuo, Y.-H., & Zhao, B. (2011). Data assimilation retrieval of electron density profiles from radio occultation measurements. *Journal of Geophysical Research, 116*, A03317. https://doi.org/10.1029/2010JA015980

138.    Yue, C., Li, W., Nishimura, Y., Zong, Q., Ma, Q., Bortnik, J., Thorne, R. M., Reeves, G. D., Spence, H. E., Kletzing, C. A., Wygant, J. R., & Nicolls, M. J. (2016). Rapid enhancement of low-energy (<100 eV) ion flux in response to interplanetary shocks based on two Van Allen Probes case studies: Implications for source regions and heating mechanisms. *Journal of Geophysical Research: Space Physics, 121*, 6430–6443. https://doi.org/10.1002/2016JA022808

139.    Zhang, J., Richardson, I. G., & Webb, D. F. (2008). Interplanetary origin of multiple-dip geomagnetic storms. *Journal of Geophysical Research, 113*, A00A12. https://doi.org/10.1029/2008JA013228

140.    Zhou, X., Tsurutani, B. T., & Gonzalez, W. D. (2000). The solar wind depletion (SWD) Event of 26 April 1999: Triggering of an Auroral "pseudobreakup" event. *Geophysical Research Letters, 27*, 4025–4028. https://doi.org/10.1029/2000GL003805

141.    Zmuda, A. J., Martin, J. H., & Heuring, F. T. (1966). Transverse magnetic disturbances at 1100 km in the auroral region. *Journal of Geophysical Research, 71*, 5033–5045. https://doi.org/10.1029/JZ071i021p05033

**Figures**

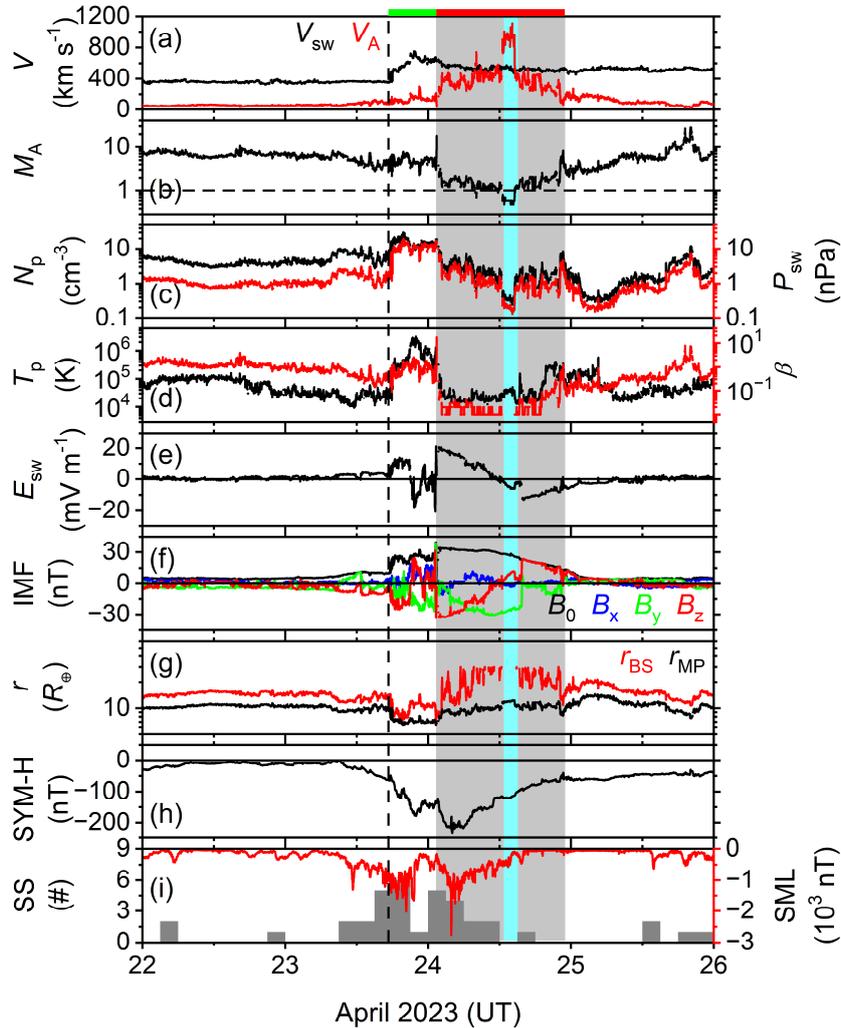

**Figure 1. Solar wind plasma and interplanetary magnetic field (IMF) during the intense storm of April 2023.** From top to bottom, the panels show: (a) solar wind flow speed $V_{sw}$ (black) and Alfvén wave speed $V_A$ (red), (b) Alfvén Mach number $M_A$, (c) proton density $N_p$ (black, legend on the left) and ram pressure $P_{sw}$ (red, legend on the right), (d) proton temperature $T_p$ (black, legend on the left) and plasma-$\beta$ (red, legend on the right), (e) electric field $E_{sw}$, (f) IMF magnitude $B_0$ (black), $B_x$ (blue), $B_y$ (green) and $B_z$ (red) components, (g) the calculated geocentric distances to the bow shock $r_{BS}$ (red) and the magnetopause $r_{MP}$ (black), (h) the geomagnetic SYM-H index, (i) number of the substorm onsets (SS, gray histograms, legend on the left) and the auroral SML index (red, legend on the right) during 22–25 April. A vertical dashed line shows a fast-forward (FF) interplanetary shock. The green bar at the top indicates an interplanetary sheath. The red bar at the top and the vertical gray shading both indicate a magnetic cloud (MC). The light-cyan shading inside the MC shows a sub-Alfvénic solar wind interval with $M_A < 1$. The horizontal dashed line in panel (b) indicates $M_A = 1$.

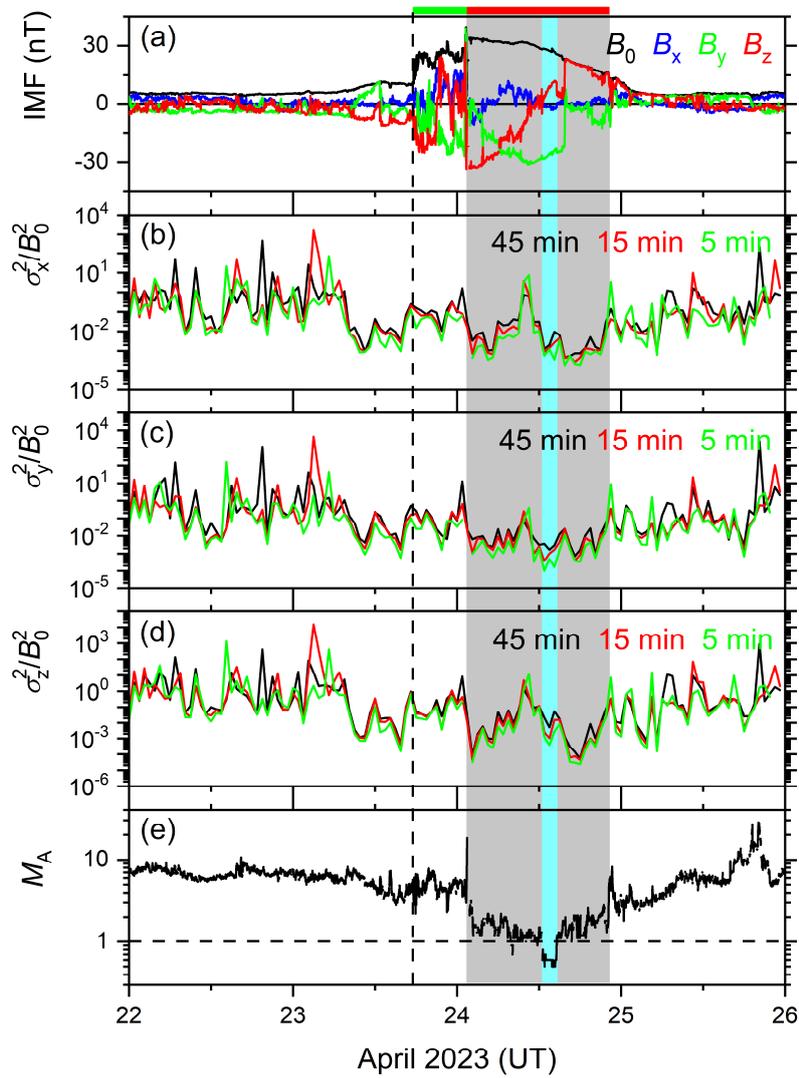

**Figure 2. Variations of the IMF during the intense storm of April 2023.** From top to bottom, the panels are: (a) the IMF $B_0$ (black) and $B_x$ (blue), $B_y$ (green), and $B_z$ (red) components, (b–d) the normalized variances of $B_x$, $B_y$, and $B_z$, and (e) $M_A$. The markings for the interplanetary FF shock, sheath, and MC are the same as in Figure 1.

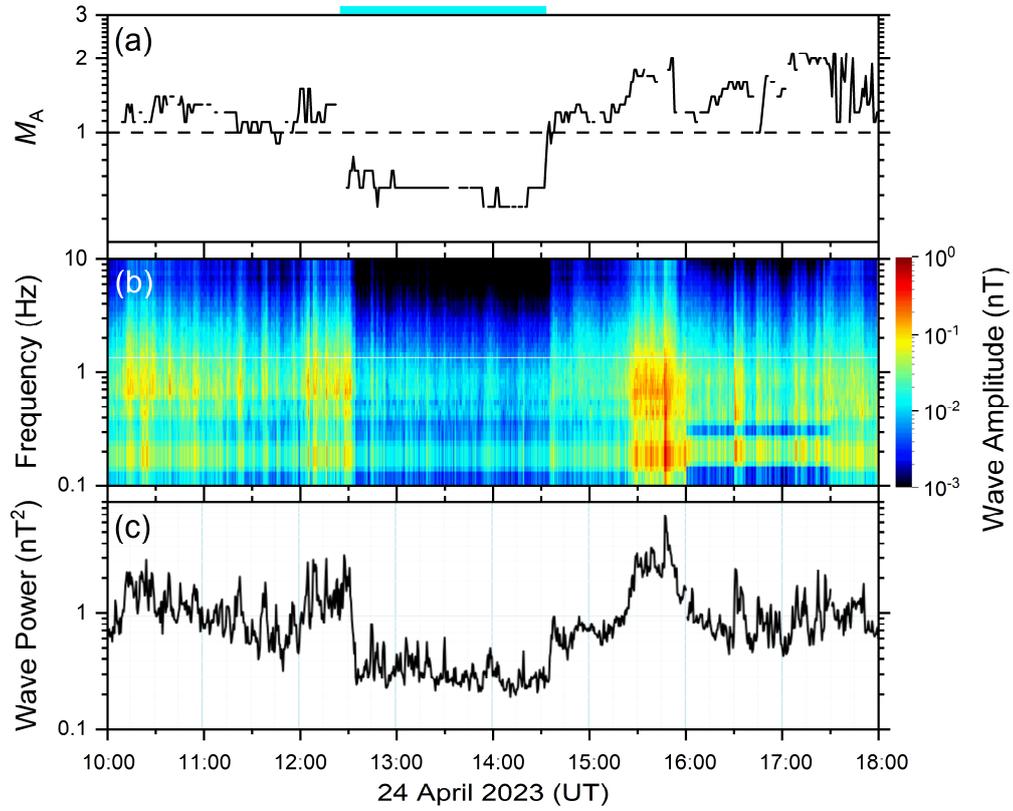

**Figure 3. Spectrogram of the magnetic field measured by the Cluster-1 spacecraft during 10:00–18:00 UT on 24 April 2023.** From top to bottom, the panels are: (a) solar wind $M_A$, (b) frequency–time spectrogram, and (c) 0.1–10 Hz wave power. The horizontal dashed line in panel (a) indicates $M_A = 1$. Color bar on the right of the panel (b) indicates the wave amplitude. Cyan bar at the top indicates the sub-Alfvénic interval.

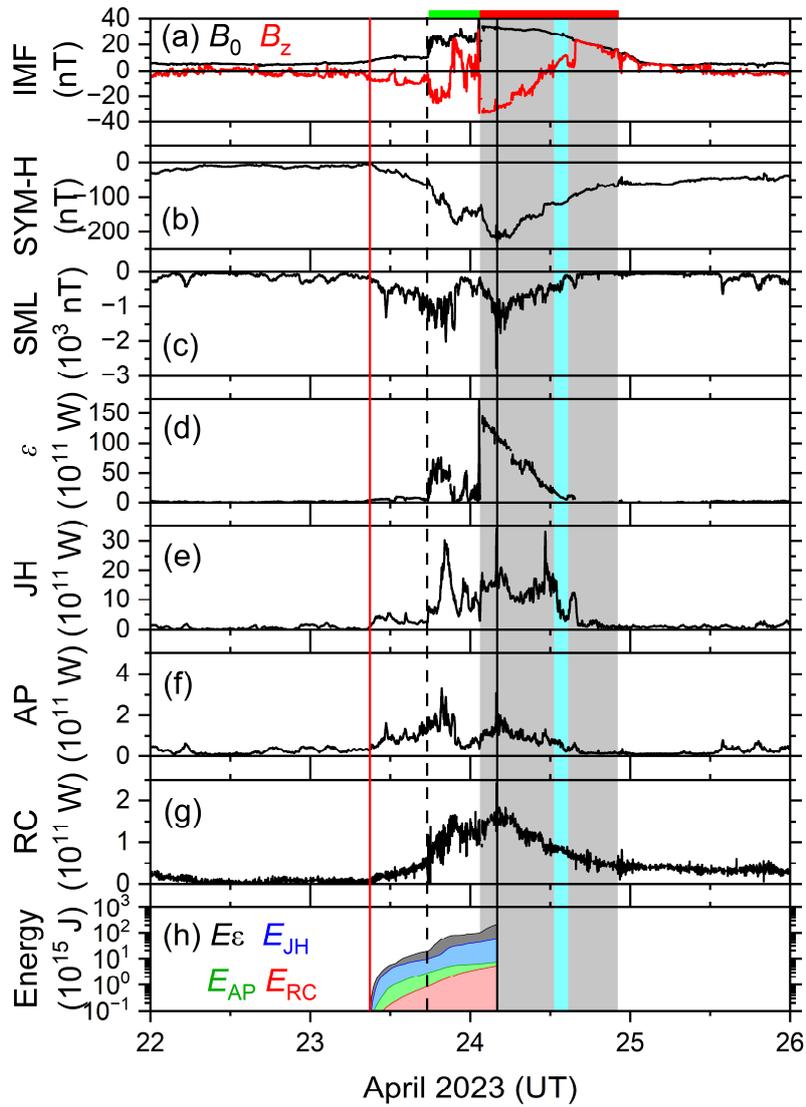

**Figure 4. Magnetosphere-ionosphere energy budget during the intense storm of April 2023.** From top to bottom, the panels are: (a) IMF $B_0$ (black) and $B_z$ component (red), (b) SYM-H, (c) SML, (d) Akasofu-$\varepsilon$ parameter, (e) Joule heating JH, (f) auroral precipitation AP, (g) ring current RC, (h) the integrated $\varepsilon$-energy input $E_\varepsilon$ (black), the integrated JH dissipation $E_{JH}$ (blue), the integrated AP dissipation $E_{AP}$ (green), and the integrated RC energy dissipation $E_{RC}$ (red). The vertical red and black solid lines indicate the storm main phase onset and termination, respectively. The markings for the interplanetary FF shock, sheath, and MC are the same as in Figure 1.

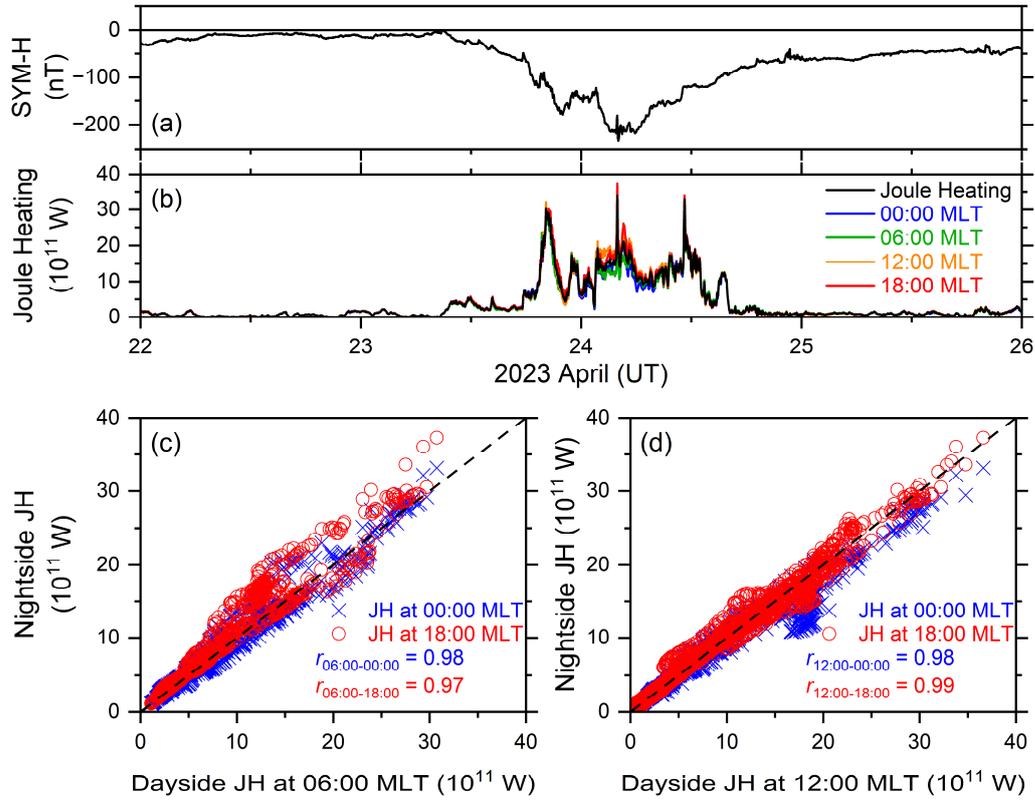

**Figure 5. Ionospheric JH during the intense storm of April 2023.** Temporal variations of (a) SYM-H, (b) the overall JH rate (black) and the JH rates at 00:00 MLT (blue), 06:00 MLT (green), 12:00 MLT (orange), and 18:00 MLT (red) time sectors; variations of the nightside JH at 00:00 MLT and 18:00 MLT sectors with (c) the dayside JH at 06:00 MLT, and (d) the dayside JH at 12:00 MLT, during the storm main phase. The corresponding linear regression coefficients (*r*) are mentioned in the panels. Diagonal (dashed) lines are added to indicate equal values.

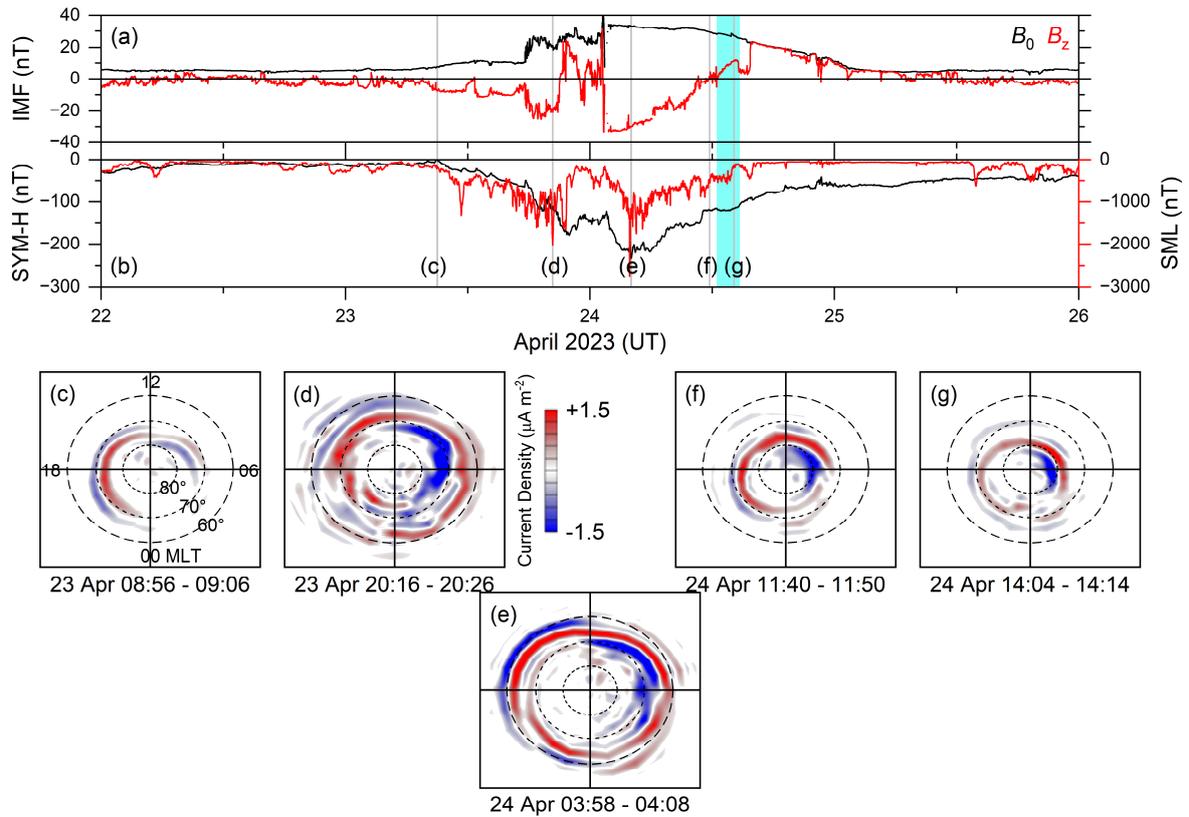

**Figure 6. Radial FACs during the intense storm of April 2023.** Temporal variations of (a) IMF $B_0$ (black) and $B_z$ (red), (b) SYM-H (black, legend on the left) and SML (red, legend on the right) indices; average radial current density for northern hemisphere plotted in AACGM and MLT coordinates during 10-minute time intervals around (c) the geomagnetic storm onset, (d) the first main phase development, (e) the second main phase SYM-H peak, (f) the recovery phase, and (g) the sub-Alfvénic wind interval during the storm recovery. The 10-minute time intervals are marked by vertical gray shadings and corresponding bottom panel numbers. Red and blue in the current density panels identify upward and downward currents, respectively. The light-cyan shading shows the sub-Alfvénic solar wind interval.

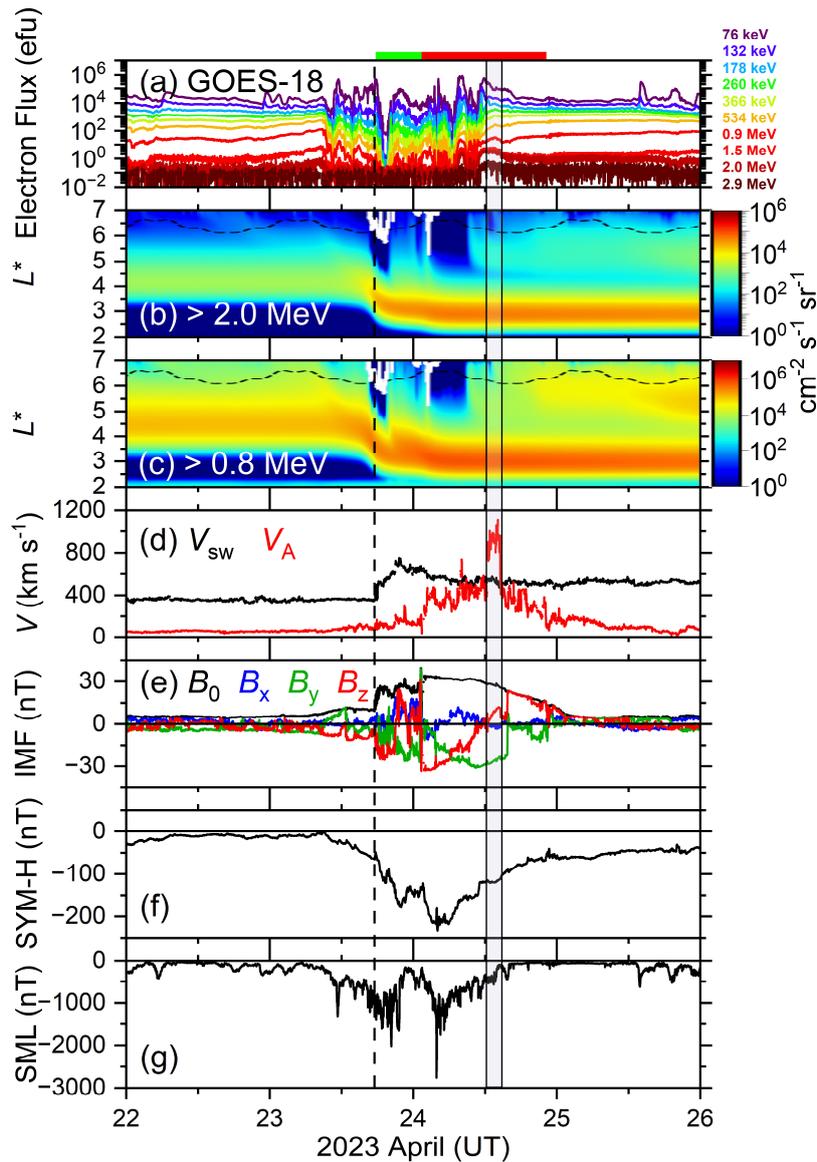

**Figure 7. The Earth's outer radiation belt during the intense storm of April 2023.** From top to bottom, panels are: (a) differential ~76 keV to 2.9 MeV electron fluxes (given in electron flux unit (efu)/cm$^{-2}$ s$^{-1}$ sr$^{-1}$) at geosynchronous orbit; $L^*$-shell variations of (b) > 2.0 MeV electron flux at ~88° pitch angle, and (c) > 0.8 MeV electron flux at ~88° pitch angle; variations of (d) $V_{sw}$ (black) and $V_A$ (red), (e) IMF magnitude $B_0$ (black), $B_x$ (blue), $B_y$ (green) and $B_z$ (red) components, (f) SYM-H index, and (g) SML index during 22–25 April 2023. The color bars at the right (in panels (b) and (c)) show fluxes of the MeV electrons. The markings for the interplanetary FF shock, sheath, and MC are the same as in Figure 1. The sub-Alfvénic interval is marked by a light-gray shading bounded by two solid vertical lines.

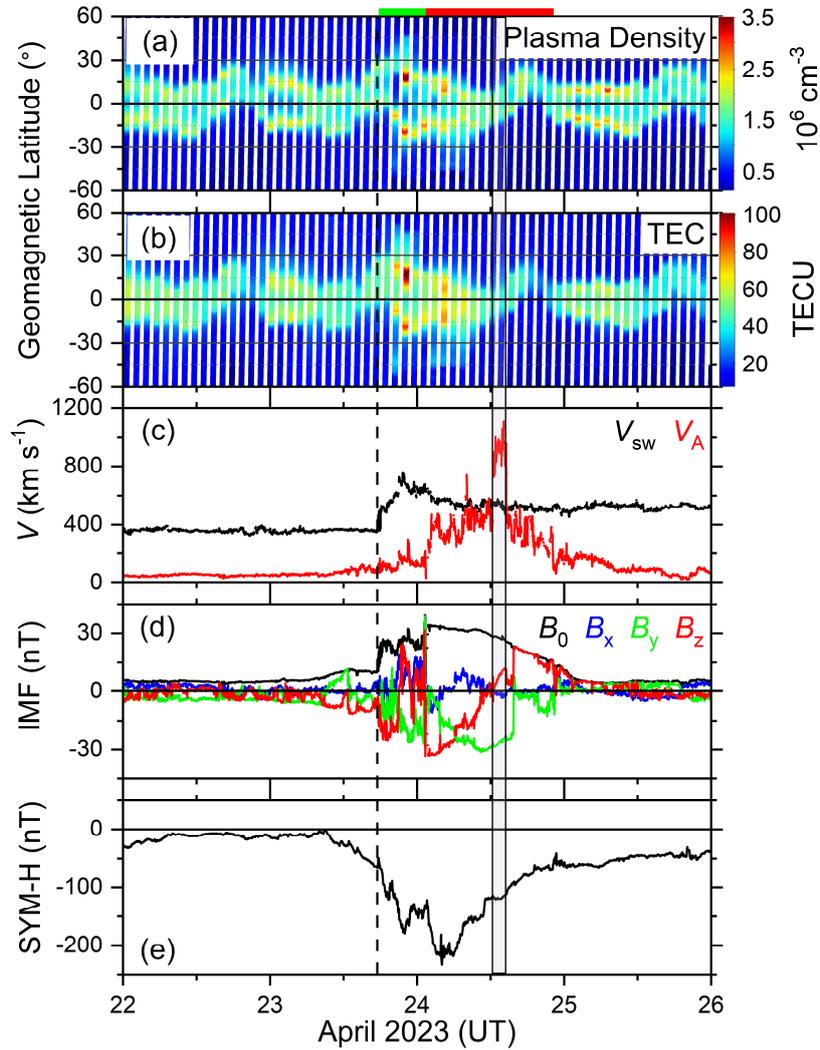

**Figure 8. Ionospheric variations during the intense storm of April 2023.** From top to bottom, panels are: (a) latitudinal distribution of ionospheric plasma density, (b) latitudinal variation of ionospheric TEC (given in TECU, 1 TECU = $10^{16}$ electrons m$^{-2}$), (c) $V_{sw}$ (black) and $V_A$ (red), (d) IMF $B_0$ (black), $B_x$ (blue), $B_y$ (green) and $B_z$ (red) components, and (e) the SYM-H index during 22–25 April. The color bars at the right (in panels (a) and (b)) show plasma density and TEC values, respectively. They correspond to the 17:00 LT passes. The markings for the interplanetary FF shock, sheath, and MC are the same as in Figure 1. The sub-Alfvénic interval is marked by a light-gray shading bounded by two solid vertical lines.